\documentclass[final,5p,times,twocolumn,fleqn]{elsarticle}
\journal{Water Research, April 2025}

\usepackage{amsthm} 
\usepackage[intlimits,fleqn]{amsmath}  
\usepackage[T1]{fontenc}
\usepackage[utf8]{inputenc}
\biboptions{sort&compress}
\usepackage{multirow}

\usepackage{bm}
\usepackage{epstopdf}
\usepackage{amssymb}
\usepackage{url}
\usepackage{enumitem} 
\usepackage{multirow}
\usepackage{hhline}
\usepackage{booktabs}
\usepackage{mathtools}
\usepackage{makecell}
\usepackage{algorithm}
\usepackage{comment}

\DeclareMathOperator*{\minimize}{minimize}
\DeclareMathOperator*{\maximize}{maximize}

\makeatother
\DeclareMathAlphabet\mathbfcal{OMS}{cmsy}{b}{n}



\usepackage{stackengine}

\newcommand{\m}{\boldsymbol}
\allowdisplaybreaks[4]
\usepackage[colorlinks = true,
linkcolor = blue,
urlcolor  = blue,
citecolor = blue,
anchorcolor = blue]{hyperref}

\usepackage[framemethod=TikZ]{mdframed}
\mdfdefinestyle{MyFrame}{%
	linecolor=black,
	outerlinewidth=1.25pt,
	roundcorner=1.25pt,
	innerrightmargin=5pt,
	innerleftmargin=5pt,}

\usepackage[noabbrev]{cleveref}

\usepackage{mathtools}

\DeclarePairedDelimiter\abs{\lvert}{\rvert}%
\DeclarePairedDelimiter\norm{\lVert}{\rVert}%

\makeatletter
\let\oldabs\abs
\def\abs{\@ifstar{\oldabs}{\oldabs*}}
\let\oldnorm\norm
\def\norm{\@ifstar{\oldnorm}{\oldnorm*}}
\makeatother

\usepackage[english]{babel}
\usepackage[utf8]{inputenc}
\usepackage[super]{nth}

\usepackage{graphicx}
\usepackage{float}
\usepackage[caption = false]{subfig}

\usepackage{array}
\usepackage{threeparttable}

\usepackage[english]{babel}
\usepackage[utf8]{inputenc}
\usepackage[super]{nth}

\usepackage{algpseudocode,float}
\usepackage{graphicx} 
\usepackage{multirow}
\usepackage{tabularx}
\usepackage{booktabs}
\usepackage{rotating, makecell}
\usepackage{hyperref}
\usepackage{tabularx}
\usepackage{makecell}
\usepackage{pifont}
\usepackage{array}
\usepackage{enumitem}
\usepackage{mathtools}
\usepackage{titlesec}
\usepackage{balance}
\titlespacing{\section}{0pt}{*0.75}{*0.75}
\titlespacing{\subsection}{0pt}{*0.3}{*0.4}
\titlespacing{\subsubsection}{0pt}{*0.4}{*0.4}

\setlist{nosep}
\usepackage{lettrine}
\usepackage{amsmath}

\usepackage{etoolbox}
\makeatletter
\patchcmd{\ps@pprintTitle}
  {Preprint submitted to}
  {In Press, }
  {}{}
\makeatother

\begin{document}
\begin{frontmatter}
\title{{\LARGE {Breaking the Flow and the Bank:} \\ {Stealthy Cyberattacks on Water Network Hydraulics}}}	
\fntext[label2]{This work is supported by the National Science Foundation under Grants 2230087 and 2404946.}
\author[inst1]{Abdallah Alalem Albustami}
\affiliation[inst1]{organization={Vanderbilt University},
addressline={Civil and Environmental Engineering Department}, 
city={Nashville},
postcode={37235}, 
state={TN},
country={US}}
\ead{abdallah.b.alalem.albustami@vanderbilt.edu}
\author[inst1]{Ahmad F. Taha}
\ead{ahmad.taha@vanderbilt.edu}

\begin{abstract}
As water distribution networks (WDNs) become increasingly connected with digital infrastructures, they face greater exposure to cyberattacks that threaten their operational integrity. Stealthy False Data Injection Attacks (SFDIAs) are particularly concerning, as they manipulate sensor data to compromise system operations while avoiding detection. While existing studies have focused on either detection methods or specific attack formulations, the relationship between attack sophistication, system knowledge requirements, and achievable impact remains unexplored. This paper presents a systematic analysis of sensor attacks against WDNs, investigating different combinations of physical constraints, state monitoring requirements, and intrusion detection evasion conditions. We propose several attack formulations that range from tailored strategies satisfying both physical and detection constraints to simpler measurement manipulations. The proposed attacks are simple and local---requiring knowledge only of targeted sensors and their hydraulic connections---making them scalable and practical. Through case studies on Net1 and Net3 benchmark networks, we demonstrate how these attacks can persistently increase operational costs and alter water flows while remaining undetected by monitoring systems for extended periods. The analysis provides utilities with insights for vulnerability assessment and motivates the development of protection strategies that combine physical and statistical security mechanisms.
\end{abstract}

\end{frontmatter}

\section{Introduction} \label{sec: intro}
Water Distribution Networks (WDNs) are undergoing a significant transformation as cities evolve towards smarter infrastructure. This change from traditional, centralized systems to more distributed intelligent networks brought tremendous improvements in real-time decision making and resource optimization. It has, nevertheless exposed vulnerabilities in cybersecurity of water systems.

In recent years, the water sector has experienced several notable cyberattacks. In 2021, a cyberattack targeted a water treatment facility in Oldsmar, Florida, attempting to poison the water supply by altering chemical levels \cite{carregaSomeoneTriedPoison2021}. More recently, in April 2024, a hacking group targeted multiple water utilities in Texas, causing a water tank to overflow and disrupting the SCADA systems that control hydraulic operations \cite{RussialinkedHackingGroup}. Despite such incidents, many cyberattacks go unnoticed or are undisclosed due to concerns over reputation and customer trust \cite{hassanzadehReviewCybersecurityIncidents2020}.

A common cyberattack on WDNs occurs when adversaries gain access to the operational technology (OT) network, typically by infiltrating the information technology (IT) network via methods like phishing, ransomware, or exploiting security vulnerabilities. Once inside the OT network, attackers can manipulate critical components such as SCADA systems, programmable logic controllers (PLCs), and sensors, allowing them to alter operational parameters \cite{aslamComprehensiveStudyCyber2023}. One particularly dangerous and common form of such attacks is a False Data Injection Attack (FDIA). In an FDIA, attackers manipulate sensor readings or actuator data to mislead state monitoring algorithms into accepting falsified values as legitimate. This is concerning in WDNs because these attacks can be \textit{stealthy}, bypassing intrusion detection (ID) methods and potentially causing significant operational and physical damage before being noticed by operators \cite{urbinaLimitingImpactStealthy2016}.
\textcolor{black}{While FDIAs have been extensively studied in fields like smart grids \cite{Liang2017, Aoufi2020} transportation networks \cite{almalki2020review, almalki2021deep}, and the control engineering literature \cite{Pang2022,Zhang2023,Zhang2024}, they have received less attention in WDNs, despite their potential to severely disrupt essential services and compromise public safety \cite{moazeniDetectionRandomFalse2022}.} The unique hydraulic characteristics of these systems necessitate tailored approaches to FDIA formulation and detection, which remain understudied. This gap motivates our work on addressing the specific challenges posed by FDIA in WDNs. The following section reviews the current state of FDIA research in this domain.

\subsection{Literature Review}
The Battle of the Attack Detection Algorithms (BATADAL) marked a significant milestone in developing and testing effective detection methods for cyber-physical attacks on WDNs \cite{taorminaBattleAttackDetection2018}. The competition featured seven algorithms for detecting cyber-physical attacks, with the only model-based approach demonstrating superior performance \cite{houshModelbasedApproachCyberphysical2018}. The attacks in BATADAL were simulated using epanetCPA, a MATLAB modeling toolbox that enables the simulation of hydraulic responses to cyber-physical attacks via EPANET \cite{taorminaToolboxAssessingImpacts2019}. As such, these attacks were not FDIAs nor were they designed with stealthiness in mind. Following BATADAL, some researchers have developed more advanced attack detection algorithms and investigated different strategies, often utilizing similar simulation environments \cite{fayzulDevelopmentEffectiveHybrid2018, ramotsoelaAttackDetectionWater2019, brentanCyberAttackDetectionWater2021, moazeniIntegratedStateestimationFramework2020}.

Despite the growing interest in WDN security, studies specifically addressing FDIAs and their stealthy variants remain limited. Urbina \textit{et al.} \cite{urbinaLimitingImpactStealthy2016} analyzed stealthy attacks in water treatment systems, finding that actuator attacks are harder to launch than sensor attacks, and that detectors using historical data (stateful) perform better than those examining individual data points (stateless). However, their work focused on pH sensors and pumps in water treatment, not addressing WDN hydraulics. Douglas \textit{et al.} \cite{douglasPressureDrivenModelingCyberPhysical2019} extended the epanetCPA toolkit to simulate cyber-physical attacks by interrupting sensor readings to impact water levels in tanks, providing insights into hydraulic impacts but not specifically formulating FDIAs. In \cite{ahmedWADIWaterDistribution2017}, a water distribution testbed (WADI) was developed using random false data injections into tank level sensors. Although their work enabled experimental assessment, the random nature of these attacks suggests they might be detectable through conventional methods such as simple ID algorithms and residual checks during state estimation (SE) processes.

More relevant to this work are the few studies that have directly addressed SFDIAs against WDN hydraulics. Ahmed \textit{\textit{et al.}} \cite{ahmedModelbasedAttackDetection2017} presented a case study on a model-based detection approach for smart WDNs. They utilized a Kalman filter for SE and compared a CUmulative SUM (CUSUM) statistic with a Bad Data Detector (BDD) for ID. They evaluated three attack scenarios, including Bias Injection Attacks (simple FDIAs that add constant offsets to measurements), Zero-Alarm Attacks (similar to SFDIAs but focused only on detection bypass), and attacks on control inputs. Their analysis of detection difficulty---showing that deceiving CUSUM is more complex than bypassing BDD due to its accumulated sum over time---aligns with findings from other domains, as their approach did not specifically address WDN hydraulics or consider physical constraints that could make attacks implausible due to mass and energy balance violations.

Exploring the offensive side of SFDIAs, Moazeni and Khazaei \cite{moazeniSequentialFalseData2021} proposed a nonlinear programming framework for modeling SFDIAs on flow rate measurements and total demand, targeting storage tanks through a bi-level optimization approach. They later adapted this strategy to target pump flow rate measurements, aiming to exceed maximum pressure heads at multiple nodes \cite{moazeniFormulatingFalseData2021}. While their work demonstrated the feasibility of attacks that satisfy hydraulic constraints while bypassing state estimation and bad data detection, it was limited to specific scenarios and detection methods. In this work, we develop a broader analysis encompassing multiple detection mechanisms and attack formulations, from random to coordinated manipulations, with varying operational targets.
Raza and Moazeni \cite{razaChanceconstrainedVulnerabilityAssessment2024} introduced a robust chance-constrained optimization strategy to identify vulnerable locations in Smart WDNs against SFDIAs, considering the probabilistic nature of water demand at junctions.

While the reviewed literature demonstrates growing attention to FDIAs in WDNs, most studies focused on specific attack scenarios or detection methods in isolation. To comprehensively understand system vulnerabilities, it was essential to analyze attacks across the full spectrum---from tailored worst-case scenarios to simpler, more common threats. This analysis required examining two critical components in cyber-physical systems: State Estimation (SE) and Intrusion Detection (ID). SE infers the system's overall state from sensor measurements, while ID analyzes discrepancies between measurements and state estimates to identify anomalies.
\textcolor{black}{A common thread across cyber-physical security literature is that effective FDIAs must evade both SE and ID mechanisms to remain stealthy. In power systems \cite{Liu2011, Yang2014} and water networks alike \cite{moazeniSequentialFalseData2021, moazeniFormulatingFalseData2021}, attackers additionally need to ensure that manipulated measurements satisfy domain-specific physical constraints---power flow equations in electrical grids and hydraulic relationships in water systems. The control engineering literature offers more generalized attack frameworks \cite{Pang2022, Zhang2023} that abstract these principles, but our approach specifically addresses the hydraulic constraints of WDNs while maintaining the essential stealth properties identified across domains. This enables practical vulnerability assessment in water infrastructure where the combination of physical laws and detection algorithms creates a multi-layered security challenge.} Although SE had been extensively researched and implemented in power grids and other cyber-physical domains, its application in pressurized WDNs remains an active area of research despite the widespread adoption of SCADA systems. This is mainly due to sparse sensor placement and a lack of accurate, calibrated models. To that end, we analyze different classes of sensor attacks, considering varying levels of system knowledge and security settings.

\subsection{Paper Contributions}
This paper investigates different classes of sensor attacks targeting WDNs. We use a Weighted Least Squares (WLS) method for state estimation (SE) and two intrusion detection (ID) methods: Cumulative Sum (CUSUM) detector and the Chi-squared detector. We formulate several SFDIA strategies with varying objectives assumptions regarding system knowledge. The main contributions of this work are summarized as follows:

\begin{itemize}[leftmargin=*]
\item Formulation and analysis of sensor attacks against WDNs by analyzing interactions between physical constraints (mass/energy balance), state estimation convergence requirements, and intrusion detection evasion. From this analysis emerges four distinct attack strategies corresponding to different attacker capabilities and system vulnerabilities, providing a foundation for understanding the relationship between an attacker's system knowledge and their ability to manipulate measurements while avoiding detection. 

\item Quantitative evaluation of how different detection approaches---dynamic CUSUM monitoring, static chi-squared tests, and physical validation checks---perform against these attacks, along with analysis of their impacts on hydraulic operations through pump scheduling and water flow management. The results highlight how local hydraulic relationships constrain attack capabilities but also enable targeted manipulations that can significantly increase operational costs and alter water flows while maintaining both physical and detection stealthiness.

\item \textcolor{black}{Development of a systematic methodology for strategic measurement selection and attack execution, including: \textit{(i)} identification of vulnerable network configurations requiring minimal sensor manipulations, \textit{(ii)} analysis of how network topology (radial vs. looped) affects attack complexity, and \textit{(iii)} creation of a comprehensive algorithm that guides practitioners through the entire attack implementation process from target selection to impact assessment. This enables utilities to systematically evaluate security vulnerabilities without requiring specialized expertise in optimization theory.}
\end{itemize}

The broader impact of this work is providing utilities with a comprehensive understanding of potential vulnerabilities across different security configurations, supporting proactive defense planning against the growing spectrum of cyber threats.
\subsection{Paper Organization}
The remainder of this paper is structured as follows: Section~\ref{sec:prelim} presents the preliminaries, including the WDN hydraulic model, SE approaches, and ID methods. Section~\ref{sec:FS-FDI} develops the full-stealth attack formulation, and Section~\ref{sec:other-attacks} presents additional attack strategies. Section~\ref{sec:case} evaluates these attacks through case studies. \textcolor{black}{Section~\ref{sec:attack_implementation} provides a systematic approach to attack design and implementation, and Section~\ref{sec:paper_limitations} discusses limitations and future research directions.} Finally, Section~\ref{sec:conc} concludes the paper.

\section{Preliminaries} \label{sec:prelim}
This section introduces the hydraulic model for the studied water distribution network, state estimation, and the intrusion detection methods utilized in this study. 

\subsection{Hydraulic Modeling of Water Distribution Networks} \label{subsec:model}
In this section, we describe the hydraulic modeling of Water Distribution Networks (WDNs), which consists of reservoirs, tanks, junctions, pipes, pumps, and valves. The hydraulic behavior of these components is governed by the principles of conservation of mass and energy. Each of these components is described below along with their governing equations. 

\subsubsection{Reservoirs}
Reservoirs in the network are considered as infinite water sources with a constant hydraulic head, representing a fixed elevation \cite{Zamzam2019}. The head at Reservoir $i$ at time step $k$ is given by: $h_i^R(k+1) = h_i^R(k)$, where $h_i^R(k)$ is the head at Reservoir $i$ at time step $k$ and remains constant over time. This assumption is valid as reservoirs are often located at a high elevation and supply water under gravity.

\subsubsection{Tanks}
Tanks are modeled with constant cross-sectional areas, and their head changes dynamically depending on the balance of inflows and outflows. The head at Tank $i$ at time step $k+1$ is given by:
\begin{equation} \label{eq:tanks}
h_i^{TK}(k+1) = h_i^{TK}(k) + \frac{\Delta t}{A_i^{TK}} \left( \sum_{j \in L_\text{in}} Q_{ij}^{\text{in}}(k) - \sum_{k \in L_\text{out}} Q_{ik}^{\text{out}}(k) \right),
\end{equation}
where $h_i^{TK}(k)$ is the head at Tank $i$ at time step $k$, $A_i^{TK}$ is the cross-sectional area of the tank, $Q_{ij}^{\text{in}}(k)$ is the flow into the tank through connected links $j$, and $Q_{ik}^{\text{out}}(k)$ is the outflow from the tank through links $k$. The head variation is directly proportional to the net water flow into or out of the tank.

\subsubsection{Junctions}
Junctions in a WDN represent intersections of water flow, where the conservation of mass ensures that the total inflows equal the total outflows plus any local demand at the junction. The mass balance at Junction $i$ is described by:
\begin{equation} \label{eq:mass_balance_model}
\sum_{j \in L_\text{in}} Q_{ij}^{\text{in}}(k) = \sum_{k \in L_\text{out}} Q_{ik}^{\text{out}}(k) + Q_i^D(k),
\end{equation}
where $Q_i^D(k)$ represents the demand at Junction $i$ at time step $k$, $Q_{ij}^{\text{in}}(k)$ is the inflow from links connected to $i$, and $Q_{ik}^{\text{out}}(k)$ is the outflow to links connected to $i$. The equation ensures that the water entering a junction is balanced by the water leaving and the demand at that junction.

\subsubsection{Pipes}
Water flowing through pipes experiences head losses due to friction and minor losses (e.g., bends). These head losses are computed using the Hazen-Williams formula, which models the relationship between flow and head loss as:
\begin{equation} \label{eq:head_loss_pipes}
\Delta h_i(k) = h_j(k) - h_k(k) = r_i \, Q_i(k) \, |Q_i(k)|^{\mu-1},
\end{equation}
where $\Delta h_i(k)$ is the head loss across Pipe $i$, $h_j(k)$ and $h_k(k)$ are the heads at Nodes $j$ and $k$, respectively, $Q_i(k)$ is the flow through the pipe, and $r_i$ is the pipe resistance coefficient, given by $r_i = \frac{4.727 \times L_i}{C_i^{1.852} \times D_i^{4.8704}}$,
where $L_i$ is the length of the pipe, $C_i$ is the roughness coefficient (typically between 100-140), and $D_i$ is the pipe diameter. The flow exponent, $\mu$, is set to 1.852, consistent with the Hazen-Williams equation \cite{linsleywater}. It is worth noting that while specific hydraulic formulas are presented here, the methods developed in this paper remain applicable to any hydraulic modeling approach as they rely only on fundamental mass and energy conservation principles.

\subsubsection{Pumps}
Pumps add energy to the water flow, increasing the head between the upstream and downstream nodes \cite{linsleywater}. The head gain provided by Pump $i$, connecting Nodes $j$ and $k$, is modeled by:
\begin{equation} \label{eq:pump}
\Delta h_i^M(k) = h_j(k) - h_k(k) = - s_i^2(k) \left( h_i^0 - \alpha_i \left(s_i^{-1}(k) Q_i^M(k)\right)^\nu \right),
\end{equation}
where $s_i(k)$ is the relative speed of the pump, $h_i^0$ is the shutoff head (the maximum head when there is no flow), $Q_i^M(k)$ is the flow through the pump, $\alpha_i$ and $\nu$ are pump-specific coefficients, and $h_j(k)$ and $h_k(k)$ are the heads at the upstream and downstream nodes, respectively. The speed $s_i(k)$ ranges between $0$ and the maximum speed $s_i^{\text{max}}$, determining the pump's operating speed.

\subsubsection{Valves}
Valves are used to control flow in the network, and in this model, they are considered as on-off components. A valve can either be fully open, allowing water to flow as in a regular pipe, or fully closed, decoupling the two connected nodes. For an open valve, the head loss across Valve $i$ connecting Nodes $j$ and $k$ is given by:
\begin{equation} \label{eq:valves}
\Delta h_i^V(k) = h_j(k) - h_k(k) = m_i Q_i^V(k) |Q_i^V(k)|,
\end{equation}
where $Q_i^V(k)$ is the flow through the valve, and $m_i$ is the minor loss coefficient associated with the valve. When the valve is closed, no flow occurs, and the two nodes are effectively decoupled.

We employ a piecewise linearization approximation for the nonlinear hydraulic components following the approach in \cite{Menke2015}. For pipes and valves, each head loss curve is segmented into linear pieces determined by connecting points calculated offline. For a pipe connecting nodes $i$ and $j$, the linearized head loss is represented through:
\begin{subequations} \label{eq:linearization}
\begin{align}
h_{j_k} - h_{i_k} - \sum_{n=1}^{N_{PW}} m_n \zeta_{n_k} - \sum_{n=1}^{N_{PW}} b_n \omega_{n_k} &= 0, \label{eq:lin_head} \\
q_{i_k} - \sum_{n=1}^{N_{PW}} \zeta_{n_k} &= 0, \label{eq:pipe_seg} \\
\sum_{n=1}^{N_{PW}} \omega_{n_k} &= 1, \label{eq:seg_select} \\
-\zeta_{n_k} + q_{n,\min}\omega_{n_k} &\leq 0, \label{eq:bound_cons1} \\
\zeta_{n_k} - q_{n,\max}\omega_{n_k} &\leq 0, \label{eq:bound_cons2}
\end{align}
\end{subequations}
where $m_n$ and $b_n$ represent the slope and intercept of segment $n$, $\zeta_{n_k}$ is the flow through segment $n$, and $\omega_{n_k}$ is a binary variable selecting segment $n$. Equation \eqref{eq:lin_head} defines the piecewise linear head loss, \eqref{eq:pipe_seg} ensures flow conservation across segments, \eqref{eq:seg_select} enforces single segment selection, \eqref{eq:bound_cons1} and \eqref{eq:bound_cons2} constrain flows within segment bounds.

For pumps, following \cite{Murguia2016}, we approximate the characteristic curves using quadratic functions:
\begin{equation}
\Delta h_{i_k}^M = \beta_1 (q_{i_k}^M)^2 + \beta_2q_{i_k}^M + \beta_3 (s_{i_k}^M)^2 + \beta_4
\end{equation}
where coefficients $\beta_1$--$\beta_4$ are determined by minimizing approximation error while ensuring convexity through $\beta_1, \beta_3 \geq 0$. This approximation maintains the relationship between pump speed, discharge, and head gain while enabling computationally tractable optimization formulations.

\subsection{Hydraulics State Estimation in WDNs}
The literature on SE for hydraulics of WDNs covers a wide range of methods. Static methods, which process measurements independently at each time step, aim to minimize residuals through iterative techniques such as sum of absolute or squared errors \cite{arsene2014, andersen_implicit_2000}. Other approaches focus on uncertainty bounds through confidence limit analysis and interval hydraulic SE \cite{bargiela_pressure_1989, vrachimis2019}. In contrast, dynamic methods like the extended Kalman filter \cite{romero-ben_nodal_2023, bartos2024} account for system evolution over time and have recently gained traction. SE provides system operators with several advantages: determining initial system state, providing real-time snapshots and detecting intrusions or anomalies \cite{tshehla_state_2017, powell_review_2000}. However, its adoption remains limited due to requirements for well-calibrated models and sufficient sensor placement for network observability. This work implements a weighted least squares (WLS) approach for state estimation, though the analysis principles remain valid for other SE methods that can provide comparable state estimates.

\subsubsection{Weighted Least Squares State Estimation}
Let $\m{x}_k \in \mathbb{R}^n$ represent the state vector at time step $k$, where $n$ is the number of state variables in the system, including flows at pipes and heads at junctions. The measurement vector is denoted by $\m{y}_k \in \mathbb{R}^m$, where $m$ represents the number of available sensor measurements, such as flow rates and pressures at selected nodes and pipes in the network. The relationship between the measurements and the system state is described by the equation $\m{y}_k = \m h(\m{x}_k) + \m{v}_k$, where $\m h(\m x) = \m H\m{x}$ is the linear measurement function with $\m H \in \mathbb{R}^{m \times n}$ being the measurement matrix, and $\m{v}_k \in \mathbb{R}^m$ represents the measurement noise, assumed to be Gaussian with zero mean and known covariance.
The objective of WLS is to minimize the weighted sum of squared residuals between the measured values $\m{y}_k$ and the predicted measurements $\m h(\hat{\m x})$ \cite{andersen_implicit_2000}. The WLS problem is formulated as:
\begin{equation} \label{eq:WLS_prob}
\minimize_{\hat{\m{x}}_k} (\m{y}_k - \m H\hat{\m{x}}_k)^\top \m{W} (\m{y}_k - \m H\hat{\m{x}}_k),
\end{equation}
which, given a linear measurement function, has the following analytical solution:
\begin{equation} \label{eq:WLS_sol}
\hat{\m{x}}_k = (\m H^\top \m W \m H)^{-1} \m H^\top \m W \m{y}_k,
\end{equation}
where $\m{W} \in \mathbb{R}^{m \times m}$ is a diagonal weight matrix, with entries representing the inverse of the measurement variances.

\textcolor{black}{While in our implementation we do not solve a formal sensor-placement optimization problem, we distribute flow and pressure sensors so that the system is observable via a WLS-based approach, ensuring the operator can reconstruct the relevant states. For measurement noise, we assume each sensor has a known variance and incorporate standard deviations directly into the weight matrix with each element representing the inverse variance of the corresponding measurement. The specific configuration of the WLS setup in this work can be found in \ref{appndx:WLS}.}

\subsection{Intrusion Detection} \label{subsec:ID}
An Intrusion Detection (ID) algorithm identifies anomalies through examining the measurement residuals, defined as the differences between sensor measurements and the state estimates, $\m{r}_k \in \mathbb{R}^m$:
\begin{equation}
\m{r}_k = \m{y}_k - \m H\hat{\m{x}}_k,
\end{equation}
Two distinct detection mechanisms are used in this study: the Cumulative Sum (CUSUM) detector and the Chi-squared ($\chi^2$) detector. 

\subsubsection{CUSUM Detection}
The Cumulative Sum (CUSUM) detector is a dynamic detection method that tracks changes in the cumulative sum of the residuals over time \cite{page1954continuous}. The detector works by accumulating deviations in the residuals from expected values, making it well-suited for identifying persistent, stealthy anomalies. While CUSUM has proven effective in various domains, its application in WDNs remains relatively limited, with a few studies applying it for water quality monitoring \cite{urbinaLimitingImpactStealthy2016} and hydraulic anomaly detection \cite{ahmedModelbasedAttackDetection2017}. 

The CUSUM detection process operates by comparing a cumulative statistic $c_k$ to a bias term $b$ and a predefined threshold $\tau$. The CUSUM procedure is defined as follows:
\begin{equation} \label{eq:CUSUM}
c_1=0, \quad c_k= \begin{cases}\max \left(0, c_{k-1}+z_k-b\right), & \text { if } c_{k-1} \leq \tau, \\ 0 \text { and } \tilde{k}=k-1, & \text { if } c_{k-1}>\tau .\end{cases}
\end{equation}
where $z_k$ represents the deviation from normal operation (distance measure), and $b \in \mathbb{R}_{> 0}$ is the bias term that adjusts the detector’s sensitivity. An alarm is triggered when $c_k$ exceeds a predefined threshold $\tau \in \mathbb{R}_{> 0}$, signaling the detection of an anomaly. Once the alarm is triggered, $c_k$ is reset to zero, and the process continues.
The CUSUM can be implemented in two ways: scalar or vectorized. The scalar approach uses $z_k = \m{r}_k^\top \Sigma^{-1} \m{r}_k$ to detect anomalies through collective evaluation of residuals, while the vectorized approach uses $z_k = |\m{r}_k|$ to monitor residuals independently, enabling detection of localized anomalies. Although \cite{Murguia2016} suggests a theoretical framework for optimal CUSUM parameter tuning, this work empirically adjusts threshold and bias parameters based on historical data to acheive the desired false alarm rate.

\subsubsection{Chi-squared ($\chi^2$) Detector} \label{sec:chi}
The Chi-squared detector is a static detection mechanism designed to identify sudden anomalies through checking whether the residual vector at time step $k$, denoted by $\m{r}_k$, falls within the expected distribution \cite{Brumback1987}. The detection process uses a chi-squared test statistic, $z_k$, defined as:
\begin{equation} \label{eq:chi}
\text{If} \;\; z_k = \m{r}_k^\top \m\Sigma^{-1} \m{r}_k > \alpha, \;\; \tilde{k}=k.
\end{equation}
where $\m\Sigma^{-1}$ is the inverse of the residual covariance matrix. An alarm is triggered when the test statistic exceeds a predefined threshold $\alpha$. This threshold is computed using the inverse regularized lower-incomplete gamma function $P^{-1}(\cdot)$, ensuring that the detection algorithm maintains a desired false alarm rate. Specifically, $\alpha$ is calculated as $\alpha = 2 P^{-1}\left(\frac{n_y}{2}, 1-\frac{1}{\gamma}\right)$, where $n_y$ is the number of independent measurements, and $\gamma$ is the desired mean time between false alarms.

Both detection methods offer complementary capabilities for identifying cyber-physical attacks. The CUSUM detector is better suited for detecting subtle, long-term deviations, and the Chi-squared detector identifies sudden anomalies that fall outside the expected range of residual values. 

\begin{figure}[t]
\centering
\includegraphics[width=\columnwidth]{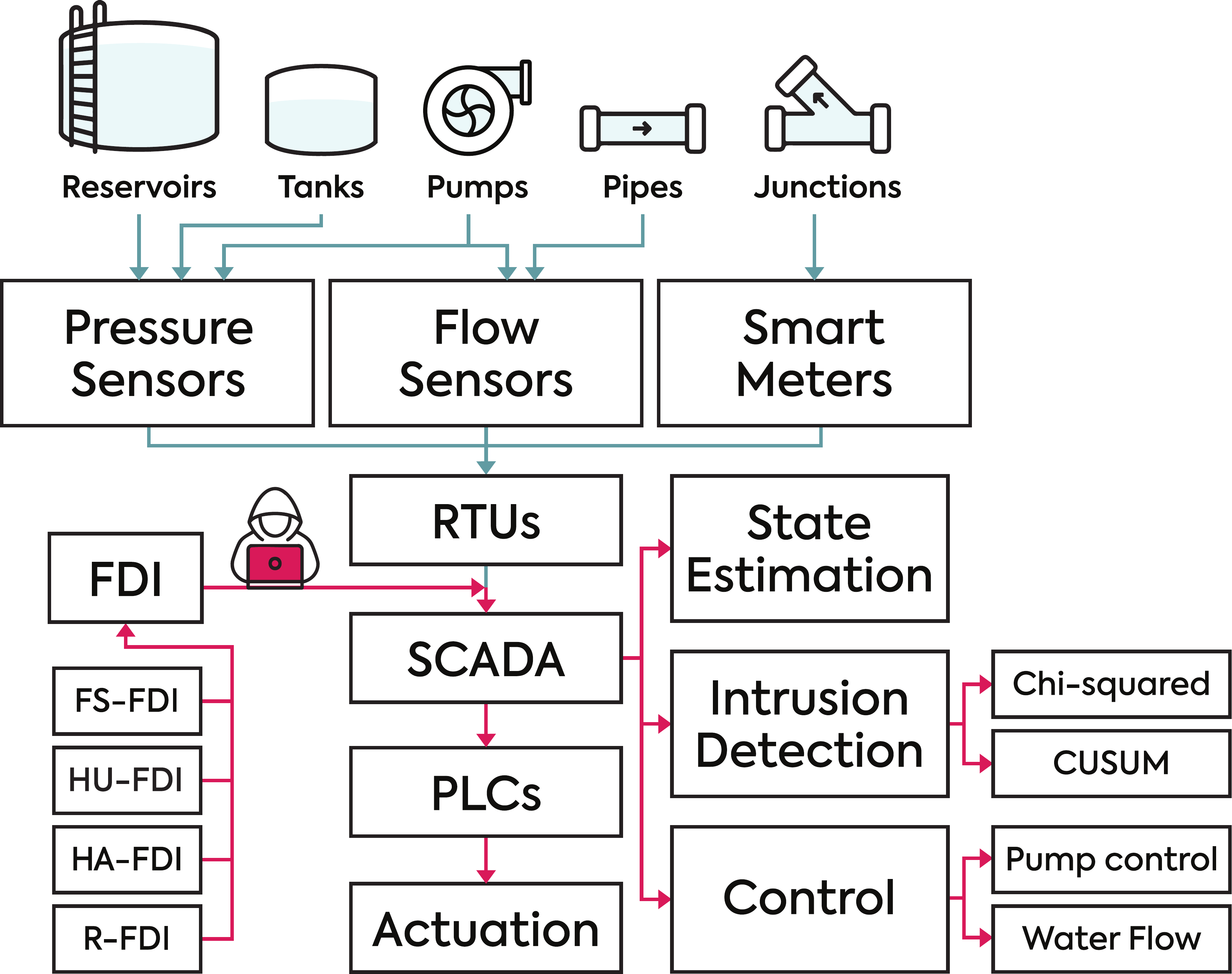}
\caption{Water distribution network architecture illustrating system components, monitoring systems, and security mechanisms. Red arrows indicate attack vectors, demonstrating how different SFDIA variants (FS-FDI: Full-Stealth FDI, HU-FDI: Hydraulics-Unaware FDI, HA-FDI: Hydraulics-Aware FDI, R-FDI: Random FDI) interact with the system's security mechanisms.}
\label{fig:Framework}
\end{figure}

Next, we present a set of tailored sensor attacks that manipulate flow, pressure, and demand measurements. The attacks are formulated with varying levels of constraint satisfaction and system knowledge requirements. Fig.~\ref{fig:Framework} illustrates the WDN architecture and its security components, showing how SFDIAs target various system layers. Tab.~\ref{tab:notations} defines the key variables and notations used throughout the analysis. Tab.~\ref{tab:sfdia_strategies} summarizes the attack strategies, comparing their characteristics and required knowledge. 

\begin{table}[t] 
\caption{Notation for Attack Formulation}
\vspace{0.3cm}
\centering
\begin{tabular}{ll}
\hline
Symbol & Description \\
\hline
$\m{x}_k \in \mathbb{R}^n$ & True state vector (flows and heads) \\
$\hat{\m{x}}_k \in \mathbb{R}^n$ & Estimated state vector \\
$\m{y}_k \in \mathbb{R}^m$ & Original measurement vector \\
$\m{y}^{a,h}_k \in \mathbb{R}^{n_h}$ & Attacked head measurements \\
$\m{y}^{a,f}_k \in \mathbb{R}^{n_f}$ & Attacked flow measurements \\
$\m{d}^a_k \in \mathbb{R}^{n_d}$ & Attacked demand measurements \\
$\m{a}^h_k \in \mathbb{R}^{n_h}$ & Head measurement attack vector \\
$\m{a}^f_k \in \mathbb{R}^{n_f}$ & Flow measurement attack vector \\
$\m{a}^d_k \in \mathbb{R}^{n_d}$ & Demand attack vector \\
$\mathbfcal{M}, \mathbfcal{E}$ & Mass and energy balance functions \\
$\mathbfcal{D}$ & Detection function \\
$\m{H}_a \in \mathbb{R}^{m \times n}$ & Measurement matrix for attacked subsystem \\
$\m{W}_a \in \mathbb{R}^{m \times m}$ & Weight matrix for attacked measurements \\
$\m{a}^{\text{drift}}_k \in \mathbb{R}^m$ & Random walk drift component \\
$\m{a}^{\text{noise}}_k \in \mathbb{R}^m$ & High-frequency noise component \\
$\m{a}^{\text{spike}}_k \in \mathbb{R}^m$ & Occasional spike component \\
\hline
\end{tabular}
\label{tab:notations}
\end{table}

\begin{table*}[t]
\centering
\caption{Comparison of FDI attack strategies highlighting implementation requirements and analytical objectives.}
\vspace{0.3cm}
\resizebox{\textwidth}{!}{%
\renewcommand{\arraystretch}{1.75}
\begin{tabular}{>{\centering\arraybackslash}m{3.2cm}>{\raggedright\arraybackslash}m{3.6cm}>{\raggedright\arraybackslash}m{4.1cm}>{\raggedright\arraybackslash}m{3.9cm}>{\raggedright\arraybackslash}m{3.3cm}}
\textbf{Attack Strategy} & \textbf{Constraint Satisfaction} & \textbf{Required Knowledge} & \textbf{Technical Description} & \textbf{Analysis Objective} \\
\hline\hline
\textbf{Full-Stealth FDI (FS-FDI) \eqref{FS-FDI}} & Mass/energy balance, SE convergence, ID bypass & 
ID parameters, local topology and hydraulic parameters, SE configuration & 
Multi-constraint optimization with coupled measurement modifications & 
Worst-case stealthy attack analysis \\
\hline
\textbf{Hydraulics-Unaware FDI (HU-FDI) \eqref{PU-CUSUM-V}, \eqref{PU-CUSUM-S}, \eqref{PU-Chi}} & SE convergence, ID bypass & 
ID parameters, SE configuration & 
Residual-based optimization with uncoupled measurements & 
Physical validation effectiveness \\
\hline
\textbf{Hydraulics-Aware FDI (HA-FDI) \eqref{HA-FDI}} & Mass/energy balance & 
Local network topology and hydraulic parameters & 
Hydraulic constraint satisfaction problem & 
Detection mechanism assessment \\
\hline
\textbf{Random FDI (R-FDI) \eqref{eq:R-FDI_base}} & None & 
Sensor operational bounds & 
Bounded measurement modifications & 
Base vulnerability metrics \\
\hline
\end{tabular}%
}
\label{tab:sfdia_strategies}
\end{table*}

\section{Full-Stealth Attack Design} \label{sec:FS-FDI}	
This section presents an optimization-based attack formulation that simultaneously satisfies physical constraints and bypasses detection mechanisms. The Full-Stealth FDI (FS-FDI) strategy is designed to: \textit{(i)} evade intrusion detection mechanisms, \textit{(ii)} maintain state estimation convergence, and \textit{(iii)} satisfy hydraulic constraints governed by mass and energy balance principles. We formally define the FS-FDI strategy through the following nonlinear optimization problem:

\begin{subequations} \label{FS-FDI}
\begin{align}
\maximize_{\m a^h_k, \m a^f_k, \m a^d_k} \quad & \sum_{i=1}^{n_h} |a^h_{i,k}| + \sum_{j=1}^{n_f} |a^f_{j,k}| + \sum_{l=1}^{n_d} |a^d_{l,k}| \label{eq:obj}\\
\text{subject to:} \quad
& \hat{\m x}^a_k = \mathbfcal{F}_a(\m y^a_k, \m W_a), \label{eq:SE} \\
& \m y^{a,h}_k = \m y^h_k + \m a^h_k, \label{eq:meas_h} \\
& \m y^{a,f}_k = \m y^f_k + \m a^f_k, \label{eq:meas_f} \\
& \m d^a_k = \m d_k + \m a^d_k, \label{eq:demand} \\
& \mathbfcal{M}(\hat{\m x}^a_k, \m d^a_k) = 0, \label{eq:mass} \\
& \mathbfcal{E}(\hat{\m x}^a_k) = 0, \label{eq:energy} \\
& \m r^a_k = [\m y^{a,h}_k; \m y^{a,f}_k] - \m h_a(\hat{\m x}^a_k), \label{eq:res} \\
& \mathbfcal{D}(\m r^a_k) \leq \beta, \label{eq:ID} \\
&  \|\hat{\m{x}}_k^a - \m{x}_{\text{ref}}^a\| \leq \epsilon, \label{eq:SE_conv} \\
& \|\m a^h_k\|_\infty \leq \alpha_h \|\m y^h_k\|_\infty, \label{eq:bounds_ah} \\
& \|\m a^f_k\|_\infty \leq \alpha_f \|\m y^f_k\|_\infty, \label{eq:bounds_af} \\
& \|\m a^d_k\|_\infty \leq \alpha_d \|\m d_k\|_\infty \label{eq:bounds_ad}
\end{align}
\end{subequations}

The objective function in \eqref{eq:obj} minimizes the sum of absolute attack values, where $\m a^h_k \in \mathbb{R}^{n_h}$, $\m a^f_k \in \mathbb{R}^{n_f}$, and $\m a^d_k \in \mathbb{R}^{n_d}$ represent attack vectors for head, flow, and demand measurements respectively, with $n_h$, $n_f$, and $n_d$ being the number of targeted sensors. This formulation promotes sparse attacks that target specific measurements, aiming to make the attack harder to detect through manual inspection or additional security measures beyond standard intrusion detection.

\textcolor{black}{The state estimation constraint \eqref{eq:SE} represents the attacker's implementation of SE over the attacked measurements and their hydraulically connected components. In the case of considering WLS, the function $\mathbfcal{F}_a$ simply represents $\hat{\m x}^a_k = (\m H^\top \m W \m H)^{-1} \m H^\top \m W \m y^a_k$. The convergence constraint \eqref{eq:SE_conv} ensures that the attacked states remain within reasonable bounds by limiting their deviation from reference states $\m x{\text{ref}}^a$ observed during normal operation. Together, these constraints guarantee that operator-side state estimation will converge under attack while requiring knowledge of only the relevant network portion. We would also like to note that in our implementation, if an operator \textit{does} employ a state estimation (SE) scheme that yields reliable estimates (so that measured values can be compared against these estimates for inrtusion detection), it does not fundamentally matter how that SE process is implemented---so long as an attacker has access to or can infer the estimates for the specific measurements they wish to manipulate. This means that regardless of whether the operator uses WLS, an Extended Kalman Filter (EKF), or any other SE approach, an attacker who can learn or approximate the relevant parameters (or simply obtain the state estimates for the sensors under attack) can design manipulations that remain undetected.} 
To implement these constraints, an attacker needs the measurement equations and weights used in the operator's SE algorithm, along with typical state bounds. Such information can be inferred from SCADA system configurations and historical operating data during reconnaissance of the OT network.

The mass balance constraint \eqref{eq:mass} enforces flow conservation at nodes using state vector components $\hat{\m x}^a_k = [\m h^a_k; \m q^a_k]$ and altered demands $\m d^a_k$ following the same flow conservation principles in \eqref{eq:mass_balance_model}. For each node $i \in \mathcal{N}$, function $\mathbfcal{M}$ ensures:
\begin{equation} \label{eq:mass_balance}
\sum_{j \in L_\text{in}} q^a_{ij_k} = \sum_{k \in L_\text{out}} q^a_{ik_k} + d^a_{i_k}
\end{equation} 
where $q^a_{ij_k}$ and $q^a_{ik_k}$ represent flows into and out of node $i$, and $d^a_{i_k}$ is the altered demand. In practice, this constraint need only be enforced within the local subnetwork affected by the attack, defined by the hydraulically connected components around targeted sensors. An attacker can identify these components through analysis of SCADA data to trace flow patterns and determine pipe connectivity.

The energy balance constraint \eqref{eq:energy} maintains consistency in the hydraulic grade line across attacked measurement paths, following the relationships established in \eqref{eq:head_loss_pipes} for pipes and \eqref{eq:pump} for pumps. For connected nodes $i$ and $j$, function $\mathbfcal{E}$ enforces:
\begin{equation} \label{eq:energy_balance}
h^a_{j_k} - h^a_{i_k} = \Delta h_{ij_k}(\m q^a_k, \m s^a_k)
\end{equation}
where $\Delta h_{ij_k}$ captures the appropriate head loss/gain based on the component type, using the piecewise linearization approach in \eqref{eq:linearization}. This constraint needs only be satisfied along paths containing attacked measurements, requiring the attacker to know pipe parameters and pump characteristics of the local subsystem.

Measurement residuals are computed in \eqref{eq:res}, where $\m h_a(\cdot)$ is a measurement function relating states to measurements. Constraint \eqref{eq:ID} ensures the attack bypasses the intrusion detection mechanism. For a CUSUM detector with vectorized distance measure, $\mathbfcal{D}(\m r^a_k) = \m c_{k-1} + |\m r^a_k| - \m b_a \leq \m \tau_a$, where the attacker only needs to know detector parameters ($b_i$ and $\tau_i$) corresponding to targeted sensors. For a chi-squared detector or CUSUM with scalar distance measure, $\mathbfcal{D}(\m r^a_k) = {\m r^a_k}^\top \m\Sigma_a^{-1} \m r^a_k \leq \alpha$, the residuals are aggregated through a weighted sum requiring knowledge of the full covariance matrix $\m\Sigma_a$ and threshold $\alpha$.

Attack magnitude constraints \eqref{eq:bounds_ah}--\eqref{eq:bounds_ad} limit attacks proportionally to original measurement magnitudes through coefficients $\alpha_h$, $\alpha_f$, and $\alpha_d$. The infinity norm ensures no single attack component exceeds these bounds.

The FS-FDI optimization problem, with piecewise linearized hydraulic constraints \eqref{eq:linearization}, takes the form of a mixed-integer linear program (MILP). While such problems are NP-hard in general, the localized nature of the attack---considers targeted components and their immediate hydraulic connections only---keeps the problem size manageable and computationally tractable regardless of the overall network size. The relatively slow dynamics of WDN operations provide sufficient time for solving the optimization at each time step using standard MILP solvers like Gurobi. Even with hydraulic time steps as small as one minute, which is more frequent than typical operational requirements, the optimization remains computationally feasible.

The implementation of the FS-FDI strategy follows Algorithm~\ref{alg:FS-FDI}, which iteratively solves the optimization problem in \eqref{FS-FDI} while ensuring both algorithmic stealthiness and physical feasibility at each time step. The algorithm requires comprehensive system knowledge including target measurements, current state estimates, detector parameters, local network topology, and hydraulic parameters. At each iteration, the algorithm computes current residuals, solves for optimal attack vectors, and verifies that both detector statistics and physical constraints remain satisfied after measurement modification. 

\begin{algorithm}[h]
\caption{Full-Stealth FDI Implementation}\label{alg:FS-FDI}
\begin{algorithmic}[1]
\Require{
measurements $\m y_k$, state estimates $\hat{\m x}_k$,  network topology and hydraulic parameters, detector parameters ($\tau, b, \alpha$), SE configuration and typical bounds}
\For{each time step $k$}
\State Compute current residuals: $\m r_k = \m y_k - \m h(\hat{\m x}_k)$
\State Initialize attack vector variables $\m a^h_k, \m a^f_k, \m a^d_k$
\State Solve optimization problem \eqref{FS-FDI}
\State Update measurements:
\State \quad $\m y^{a,h}_k = \m y^h_k + \m a^h_k$
\State \quad $\m y^{a,f}_k = \m y^f_k + \m a^f_k$
\State \quad $\m d^a_k = \m d_k + \m a^d_k$
\State Verify detector statistics remain within thresholds
\State Verify physical constraints are satisfied
\EndFor
\State \Return modified measurements $[\m y^{a,h}_k; \m y^{a,f}_k]$ and demands $\m d^a_k$
\end{algorithmic}
\end{algorithm}

The FS-FDI strategy provides a general framework for analyzing worst-case stealthy sensor attacks against WDN hydraulics. The formulation ensures attack stealthiness through: \textit{(i)} intrusion detection bypass, \textit{(ii)} state estimation convergence, and \textit{(iii)} physical constraint satisfaction. 
While the strategy requires significant system knowledge (see Tab.~\ref{tab:sfdia_strategies}), it aligns with the literature on worst-case SFDIA formulation for cyber-physical systems. The proposed formulation serves several purposes. First, it provides a benchmark for evaluating system resilience against stealthy and tailored worst case sensor attacks, which are becoming increasingly common in critical infrastructure. Second, it enables operators to systematically identify vulnerabilities in their monitoring systems and develop targeted protection strategies. Third, despite its apparent complexity, the framework's reliance on standard hydraulic principles and common operational data makes it practically implementable for both security testing and attack simulation. The larger time scales associated with WDN hydraulics provide flexibility for implementing these optimization-based attacks within feasible operational intervals.

While FS-FDI provides a complete attack formulation satisfying all security constraints, we next examine strategies that relax different combinations of these constraints to analyze how different constraints affect attack capabilities and impact.

\section{Constraint-Relaxed Attack Strategies} \label{sec:other-attacks}
This section presents three attack formulations with varying degrees of sophistication, examining how different constraint relaxations affect attack capabilities and required system knowledge.
\subsection{Hydraulics-Unaware FDI (HU-FDI)} \label{HU-FDI}
A Hydraulics-Unaware FDI attack is designed to maintain algorithmic stealthiness through ID bypass and SE convergence while neglecting the physical consistency requirements. The strategy enables analysis of how hydraulic constraints influence attack feasibility and helps quantify the trade-off between attack complexity and required system knowledge. Two main approaches are typically used in designing HU-FDI attacks, an optimization-based approach and a closed-form solution \cite{Urbina2016}.

For the optimization approach, the HU-FDI strategy modifies the formulation in \eqref{FS-FDI} by removing the hydraulic constraints \eqref{eq:mass} and \eqref{eq:energy}, yielding a computationally simpler formulation that requires only detection parameters and targeted measurement information. 

We also derive closed-form solutions for attack vectors that maintain detector statistics at their respective thresholds for both detection mechanisms presented in Section~\ref{subsec:ID}. These solutions take different forms depending on how the detector evaluates the deviation between measurements and estimates---either by examining each residual independently or by considering their collective weighted sum. For the vectorized CUSUM implementation where $z_k=|\m{r}_k|$, the attack vector components are designed independently \cite{Quinonez2020, Urbina2016}:
\begin{equation} \label{PU-CUSUM-V}
a_{k,i} = \begin{cases}
\pm(\tau_i+b_i-c_{k-1,i})-r_{k,i}, & \text{if } k=k^* \\
b_i-r_{k,i}, & \text{if } k>k^*
\end{cases}
\end{equation}
where $i \in {1,\ldots,m}$ indexes the targeted measurements. Here, the attacker only needs knowledge of individual residual parameters and measurements.
For scalar distance measures, the closed-form solutions require more comprehensive system knowledge. The CUSUM detector with a scalar distance measure, $z = \m r_k^\top \m \Sigma^{-1} \m r_k$, yields \cite{Murguia2016}:
\begin{equation} \label{PU-CUSUM-S}
\m a_k = \begin{cases}
\m \Sigma^{\frac{1}{2}} \m \Gamma\left(\sqrt{\frac{\tau+b-c_{k-1}}{n}}, \ldots, \sqrt{\frac{\tau+b-c_{k-1}}{n}}\right)^\top-\m r_k, & \text{if } k=k^* \\
\m{\Sigma}^{\frac{1}{2}} \m{\Gamma}\left(\sqrt{\frac{b}{n}}, \ldots, \sqrt{\frac{b}{n}}\right)^\top-\m r_k, & \text{if } k>k^*
\end{cases}
\end{equation}
where $\m{\Gamma}$ represents the selection matrix for targeted measurements. Similarly, for the chi-squared detector \cite{Murguia2016}:
\begin{equation} \label{PU-Chi}
\m{a}_k = \m{\Sigma}^{\frac{1}{2}} \m{\Gamma}\left(\sqrt{\frac{\alpha}{n}}, \ldots, \sqrt{\frac{\alpha}{n}}\right)^\top-\m r_k,
\end{equation}
For an ID mechanism that employs a scalar distance measure, an attacker implementing the closed-form solution would need comprehensive knowledge of the entire measurement vector, state estimates, residual covariance matrix, and detector parameters. This requirement stems from the scalar nature of the detection statistic, where residuals are collectively evaluated. In contrast, an optimization-based approach provides a more practical alternative, requiring only knowledge of the targeted measurements and their corresponding detector parameters. This makes the optimization-based approach more suitable for real-world implementation against scalar detection schemes. 

While HU-FDI attacks can successfully evade detection, the manipulated measurements and resulting state estimates could violate physical laws, potentially alerting operators through obvious deviations from expected hydraulic behavior. This highlights the importance of incorporating physical constraints in attack design, as demonstrated by the FS-FDI strategy.

\subsection{Hydraulics-Aware FDI (HA-FDI)}
A Hydraulics-Aware FDI strategy represents attacks on systems where operators rely primarily on physical validation rather than intrusion detection mechanisms. This scenario is particularly relevant for WDNs where SE and ID systems may not be implemented due to cost or complexity constraints \cite{Tshehla2017State}. 

The HA-FDI strategy is a simplified version of the FS-FDI approach. It modifies the FS-FDI formulation in \eqref{FS-FDI} by removing the ID and SE-related constraints in \eqref{eq:res}--\eqref{eq:SE_conv}. By doing so, it focuses on satisfying physical constraints to ensure that the manipulated measurements remain hydraulically plausible. This approach allows the attacker to construct attack vectors that evade physical validation checks while requiring only knowledge of the local network topology and hydraulic parameters. The HA-FDI optimization is expressed as follows:
\begin{subequations} \label{HA-FDI}
\begin{align}
\maximize_{\m a^h_k, \m a^f_k, \m a^d_k} \quad & \sum_{i=1}^{n_h} |a^h_{i,k}| + \sum_{j=1}^{n_f} |a^f_{j,k}| + \sum_{l=1}^{n_d} |a^d_{l,k}| \label{eq:pa_obj}\\
\text{subject to:} \quad
& \m y^{a,h}_k = \m y^h_k + \m a^h_k, \label{eq:pa_meas_h} \\
& \m y^{a,f}_k = \m y^f_k + \m a^f_k, \label{eq:pa_meas_f} \\
& \m d^a_k = \m d_k + \m a^d_k, \label{eq:pa_demand} \\
& \mathbfcal{M}(\hat{\m x}^a_k, \m d^a_k) = 0, \label{eq:pa_mass} \\
& \mathbfcal{E}(\hat{\m x}^a_k) = 0, \label{eq:pa_energy} \\
& \|\m a^h_k\|_\infty \leq \alpha_h \|\m y^h_k\|_\infty, \label{eq:pa_bounds_ah} \\
& \|\m a^f_k\|_\infty \leq \alpha_f \|\m y^f_k\|_\infty, \label{eq:pa_bounds_af} \\
& \|\m a^d_k\|_\infty \leq \alpha_d \|\m d_k\|_\infty. \label{eq:pa_bounds_ad}
\end{align}
\end{subequations}

This strategy provides a benchmark for assessing the limitations of detection systems that rely solely on physical validation checks. It highlights how attacks can bypass these checks by exploiting the hydraulic properties of the network. In practice, the HA-FDI strategy requires less computational effort compared to FS-FDI, as it avoids the need to account for ID or SE mechanisms. However, this also means that HA-FDI attacks may result in less algorithmically stealthy manipulations, potentially raising suspicion in systems equipped with detection systems.

\subsection{Random FDI (R-FDI)}
The Random FDI strategy represents attacks executed with minimal system knowledge and complexity, as a baseline for evaluating the effectiveness of both detection mechanisms and more advanced attack strategies. R-FDI implements structured randomization that reflects attack patterns that might emerge from automated scripts or basic manipulation tools.

We define the R-FDI attack through a multi-pattern random process that combines three attack components:
\begin{equation} \label{eq:R-FDI_base}
a_{i,k} = \begin{cases}
a_{i,k}^{\text{drift}} + a_{i,k}^{\text{noise}} + a_{i,k}^{\text{spike}}, & \text{if } i \in \mathcal{T} \text{ and } k \geq k^* \\
0, & \text{otherwise}
\end{cases}
\end{equation}

The drift component introduces systematic bias through a random walk:
\begin{equation} \label{eq:R-FDI_drift}
a_{i,k}^{\text{drift}} = a_{i,k-1}^{\text{drift}} + \delta_{i,k}, \quad \delta_{i,k} \sim \mathcal{N}(0, \sigma_d^2)
\end{equation}

The noise component adds high-frequency perturbations:
\begin{equation} \label{eq:R-FDI_noise}
a_{i,k}^{\text{noise}} = \nu_{i,k} \cdot \alpha_n \cdot y_{i,k}, \quad \nu_{i,k} \sim \mathcal{N}(0, \sigma_n^2)
\end{equation}

The spike component introduces occasional large deviations:
\begin{equation} \label{eq:R-FDI_spike}
a_{i,k}^{\text{spike}} = \begin{cases}
s_{i,k} \cdot \alpha_s \cdot y_{i,k}, & \text{if } u_{i,k} \leq p_s \\
0, & \text{otherwise}
\end{cases}
\end{equation}
where $s_{i,k} \sim \mathcal{U}(-1,1)$ generates random spike magnitudes, $u_{i,k} \sim \mathcal{U}(0,1)$ determines spike occurrence with probability $p_s$, and $\alpha_s$ scales spike magnitude.

The final attack magnitude is constrained to maintain basic operational plausibility through:
\begin{equation} \label{eq:R-FDI_bound}
a_{i,k}^{\text{final}} = \begin{cases}
\alpha_{\max}|y_{i,k}|, & \text{if } a_{i,k} > \alpha_{\max}|y_{i,k}| \\
-\alpha_{\max}|y_{i,k}|, & \text{if } a_{i,k} < -\alpha_{\max}|y_{i,k}| \\
a_{i,k}, & \text{otherwise}
\end{cases}
\end{equation}

\begin{figure}[t] 
\centering
\includegraphics[width=\columnwidth]{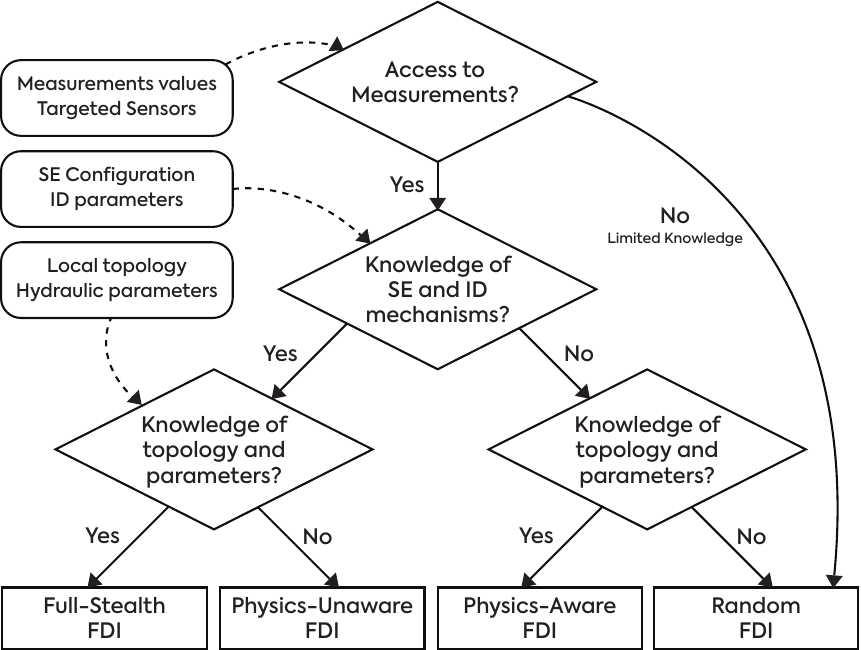}
\caption{Decision-making flowchart for selecting FDI attack strategies based on available knowledge}
\label{fig:strategies}
\end{figure}

The implementation follows Algorithm~\ref{alg:R-FDI}. Fig.~\ref{fig:strategies} presents a decision flowchart that guides attack strategy selection based on three key knowledge components: measurement access, understanding of security mechanisms (SE/ID), and network topology/parameters.

\begin{algorithm}[h]
\caption{Random FDI Implementation}\label{alg:R-FDI}
\begin{algorithmic}[1]
\Require{Measurements $\m y_k$, Target set $\mathcal{T}$, Parameters $\alpha_{\max}, \sigma_d, \sigma_n, \alpha_n, \alpha_s, p_s$}
\State Initialize: $\m a_0^{\text{drift}} = \m 0$
\For{each time step $k$}
\For{each target $i \in \mathcal{T}$}
\State Generate $\delta_{i,k} \sim \mathcal{N}(0, \sigma_d^2)$
\State Update drift: $a_{i,k}^{\text{drift}} = a_{i,k-1}^{\text{drift}} + \delta_{i,k}$
\State Generate noise: $\nu_{i,k} \sim \mathcal{N}(0, \sigma_n^2)$
\State Compute noise: $a_{i,k}^{\text{noise}} = \nu_{i,k} \cdot \alpha_n \cdot y_{i,k}$
\State Generate $u_{i,k} \sim \mathcal{U}(0,1)$
\If{$u_{i,k} \leq p_s$}
\State Generate $s_{i,k} \sim \mathcal{U}(-1,1)$
\State Compute spike: $a_{i,k}^{\text{spike}} = s_{i,k} \cdot \alpha_s \cdot y_{i,k}$
\Else
\State Set $a_{i,k}^{\text{spike}} = 0$
\EndIf 
\State Combine components via \eqref{eq:R-FDI_base}
\State Apply bounds via \eqref{eq:R-FDI_bound}
\State Update measurement: $y_{i,k}^a = y_{i,k} + a_{i,k}^{\text{final}}$
\EndFor
\EndFor
\State \Return Modified measurements $\m y^a_k$
\end{algorithmic}
\end{algorithm}

\subsection{Impact on Hydraulic Operations}
We evaluate the proposed attacks' impact on two fundamental hydraulic operations: optimal pump scheduling and water flow.  \textcolor{black}{In the hydraulic model considered here, the overall system state $\m x \in \mathbb{R}^n$ comprises the following physical quantities:
\[
\m{x} = \begin{bmatrix} \m{h}_j^\top & \m{h}_t^\top & \m{q}_p^\top & \m{q}_{\text{pump}}^\top & \m{s}_{\text{pump}}^\top \end{bmatrix}^\top \in \mathbb{R}^{n_j + n_t + n_p + 2n_m},
\]
where $\m{h}_j \in \mathbb{R}^{n_j}$ represents the hydraulic heads at junctions (in ft), $\m{h}_t \in \mathbb{R}^{n_t}$ represents the hydraulic heads at tanks (in ft), $\m{q}_p \in \mathbb{R}^{n_p}$ represents the flow rates in pipes (in GPM), $\m{q}_{\text{pump}} \in \mathbb{R}^{n_m}$ represents the flow rates through pumps (in GPM), and $\m{s}_{\text{pump}} \in \mathbb{R}^{n_m}$ represents the pump speeds (dimensionless, expressed as a fraction of maximum speed).}

The pump scheduling optimization minimizes operational costs while maintaining hydraulic constraints through the following optimization:
\begin{subequations} \label{eq:pump_control}
\begin{align}
\minimize_{\m{s}, \m{q}} \quad & \sum_{i=1}^{N_p} \varphi_{EL} \frac{\rho_w g}{\eta_i} q_i \Delta h_i^M \label{eq:pump_obj} \\
\text{subject to:} \quad
& \text{Mass balance \eqref{eq:mass_balance_model}} \label{eq:pump_mass} \\
& \text{Energy balance \eqref{eq:head_loss_pipes}, \eqref{eq:pump}} \label{eq:pump_energy} \\
& \m{s} \in [0, \m{s}_{\max}], \quad \m{q} \in [\m{q}_{\min}, \m{q}_{\max}] \label{eq:pump_bounds}
\end{align}
\end{subequations}
where $\varphi_{EL}$ represents electricity price ($\$/\text{kWh}$), $\rho_w$ is water density ($\text{kg}/\text{m}^3$), $g$ is gravitational acceleration ($\text{m}/\text{s}^2$), $\eta_i$ is pump efficiency ($\%$), $q_i$ is flow rate through pump $i$ ($\text{m}^3/\text{s}$), and $\Delta h_i^M$ is the head gain across pump $i$ ($\text{m}$).

The water flow problem verifies physical consistency by removing the cost minimization objective \eqref{eq:pump_obj} while maintaining feasibility constraints \eqref{eq:pump_mass}--\eqref{eq:pump_energy}. These operations represent typical SCADA functionalities in modern WDNs, where operators rely on automated systems for control decisions and hydraulic validation.
By manipulating sensor measurements while maintaining apparent physical consistency, the proposed attacks can induce suboptimal pump schedules or false feasibility assessments. Next, we evaluate these impacts through case studies on standard benchmark networks.

\section{Case Studies}\label{sec:case}
\begin{figure}[t]
\centering
\includegraphics[width=1\columnwidth]{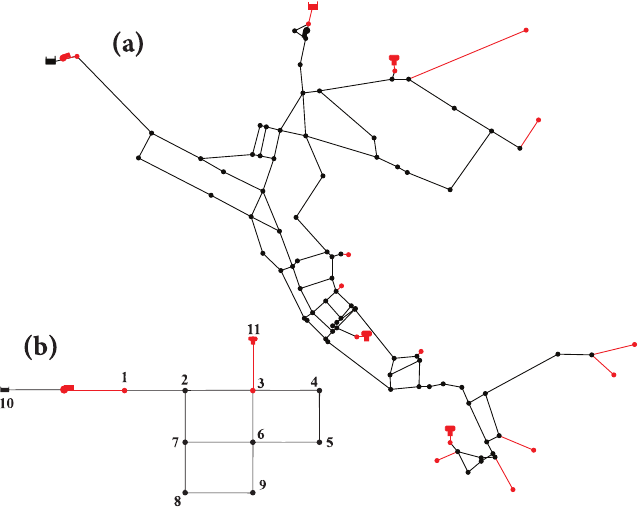}
\caption{Studied Water Distribution Networks: (a) Net3, (b) Net1}
\label{fig:networks}
\end{figure}

This section evaluates the proposed strategies on two standard EPANET benchmark networks: Net1 and Net3 \cite{rossman2020epanet}. The schematic layouts of these networks are shown in Fig.~\ref{fig:networks}. All system parameters are extracted using the EPANET toolbox on MATLAB (R2024a), with optimization solved using Gurobi 11.0.2. 
Intrusion detection employs a vectorized CUSUM statistic, with parameters tuned using historical data during normal operation. The detection threshold (\(\tau\)) is set as the mean of residuals plus three times the standard deviation, while the parameter (\(b\)) is selected as the mean of residuals plus 0.5 times the standard deviation. For the chi-squared, detector, the threshold is tuned according to Section~\ref{sec:chi}. The electricity cost is assumed to be \$0.175 per kWh. \textcolor{black}{For Net1, the WLS relies on 5 sensors across the network, providing approximately 24\% coverage of all 21 potential measurement points (12 pipe flows and 9 head measurements). The sensor configuration includes 4 flow sensors monitoring pipes 1, 5, 7, and 9, and a pressure sensors at junction 3.
Additionally, we assume that the operator measures the water level at the tank, the pump flow rate, and the demand at Junctions. In the case of Net1, Junction 1 is the only demand junction. The specific placement of sensors for Net3 is omitted for brevity.
The full implementation, including all attack formulations, simulation framework, and examples, will be released on GitHub \cite{codes}.}

Prior to incorporating sensor measurements into SCADA operations (SE, ID, pump control, and water flow) the system performs hydraulic validation checks to ensure physical consistency. These checks verify that received measurements satisfy mass and energy balance constraints across the network. This step is what necessitates that any \textit{successful} sensor manipulation must maintain physical plausibility.

Through these case studies, we investigate: \textit{(i)} How effectively can attacks bypass both detection and physical validation? \textit{(ii)} What impact can they achieve on system operations? \textit{(iii)} Do these approaches remain viable for larger networks?

The following analyses are all conducted on the Net1 network, with Section \ref{subsec:net3} extending to Net3 to demonstrate the scalability of the proposed strategies.

\subsection{Full-Stealth FDI Analysis}
For the Full-Stealth FDI evaluation, we target three critical measurements: Pump 1 flow sensor, the demand meter at Junction 1, and Pipe 1 flow sensor. The attack magnitude is constrained to 10\% of the true measurements. Network observability is ensured through strategic sensor placement: 7 flow sensors across 12 pipes and 6 pressure sensors among 9 junctions, providing 60\% coverage of potential measurement points. Simulations are conducted over a 24-hour horizon, with attacks initiated at \(t = 11\) and terminated at \(t = 20\).

\begin{figure}[http]
\centering
\includegraphics[width=\columnwidth]{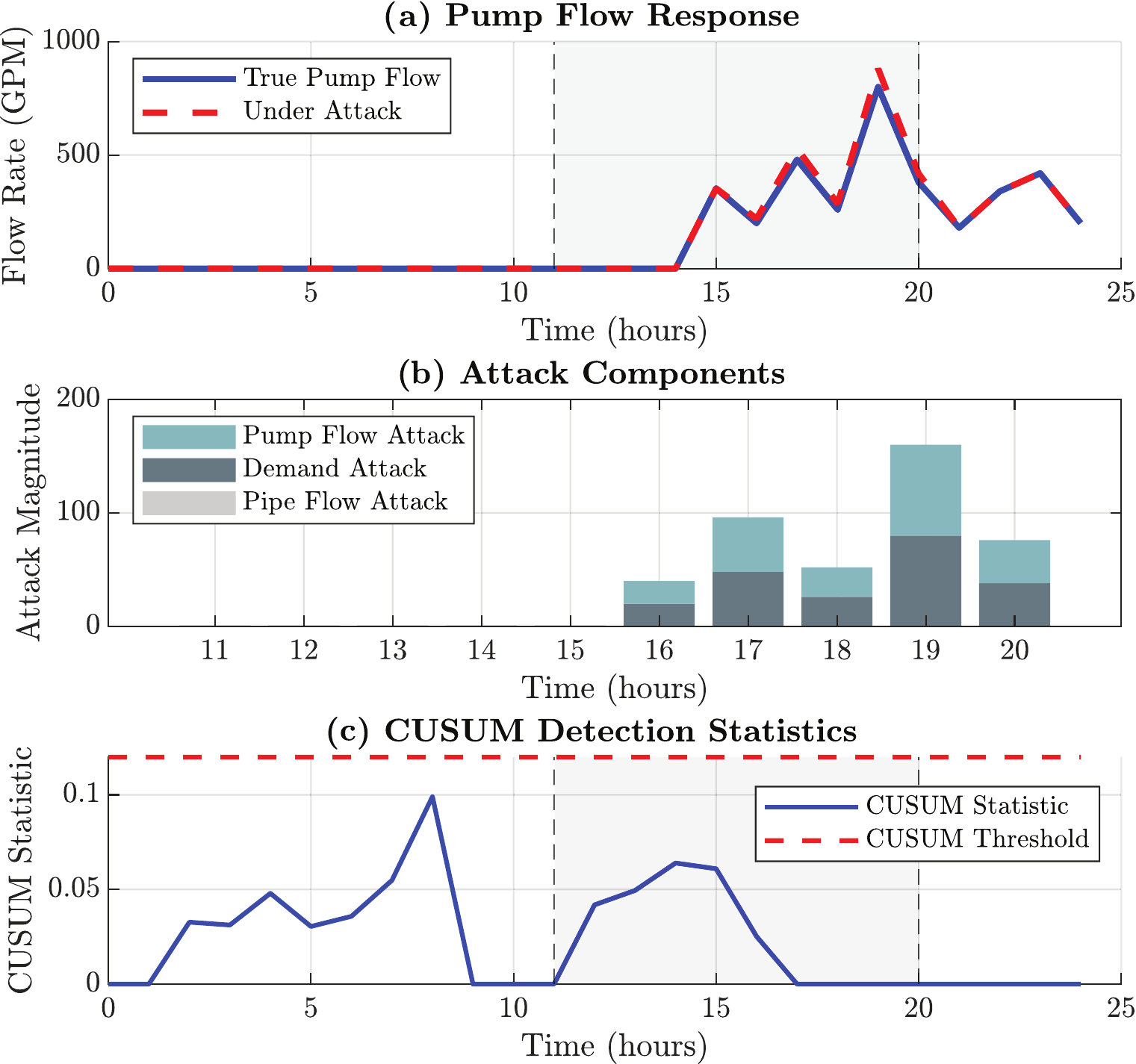}
\caption{Full-Stealth FDI attack demonstrating stealthy manipulation of pump operations while maintaining detection avoidance}
\label{fig:FS}
\end{figure}

The attack's success depends on its coordinated manipulation of multiple sensors---when the pump flow measurement is altered, corresponding changes are made to connected pipe flows and junction demands to maintain hydraulic consistency. The optimization objective maximizes the magnitude of modifications across accessible measurements while searching for a feasible solution that satisfies both physical and detection constraints. The attacker's ability to construct such a solution critically depends on having access to a sufficient set of hydraulically coupled measurements that can be manipulated simultaneously. This coordination allows the attack to induce suboptimal pump operations while ensuring all manipulated measurements appear physically plausible to operators. Fig.~\ref{fig:FS} demonstrates the impact of the FS-FDI strategy \eqref{FS-FDI} on the optimal pump control problem \eqref{eq:pump_control}. Between hours 11 and 15, the optimization was unable to find a solution that could increase pump flow or demand while maintaining stealthiness against intrusion detection and physical plausibility constraints. This is likely due to the pump's inactive status during these hours, where any sudden changes would trigger CUSUM detection. However, from hours 15 to 20, the optimization successfully identified opportunities to manipulate both the demand meter at Junction 1 and pump flow sensor, effectively increasing pumping costs while bypassing both physical validation checks and intrusion detection systems.

\subsection{Constraint-Relaxed FDI Analysis}
\begin{figure}[http]
\centering
\includegraphics[width=\columnwidth]{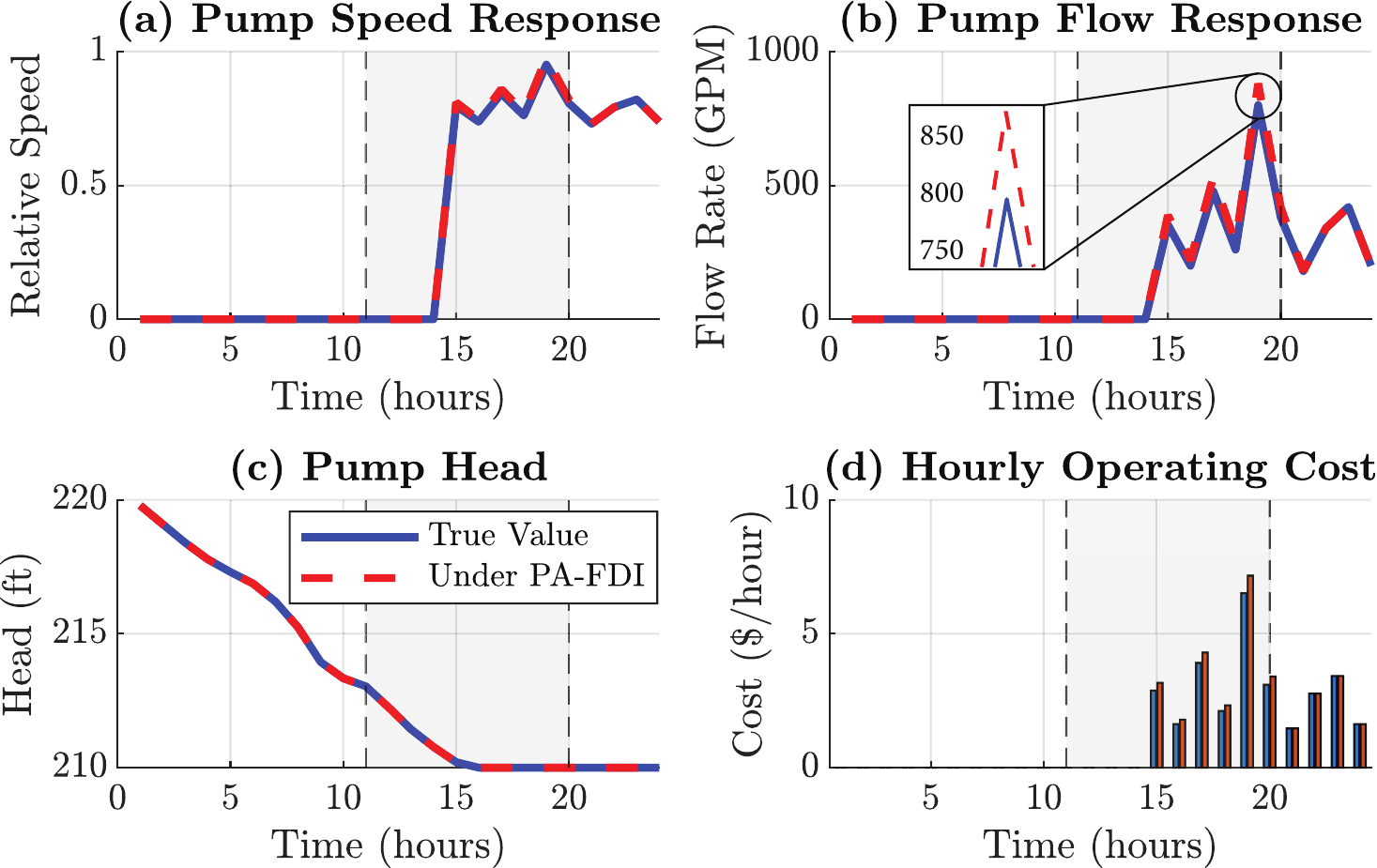}
\caption{Hydraulics-Aware FDI impacts on pump operations and associated costs}
\label{fig:PA_pump_control}
\end{figure}

Next, we analyze the Hydraulics-Aware FDI strategy under identical timing and targeting parameters. This attack achieved marginally higher impact on pump operations compared to the FS-FDI, suggesting that physical consistency in this scenario posed more restricting constraints than intrusion detection requirements. Fig.~\ref{fig:PA_pump_control} demonstrates these effects when maintaining only hydraulic consistency without consideration for detection avoidance.

The relative speed and flow responses in Fig.~\ref{fig:PA_pump_control}(a,b) show that despite having more degrees of freedom by ignoring detection constraints, the attack's impact remains bounded by the need to satisfy mass and energy balance equations, as evidenced by the preserved head patterns in Fig.~\ref{fig:PA_pump_control}(c). This relatively constrained impact is also confirmed in Fig.~\ref{fig:comparison}, where HA-FDI shows only modest additional cost increases compared to FS-FDI across the full attack duration.

\begin{figure}[t]
\centering
\includegraphics[width=\columnwidth]{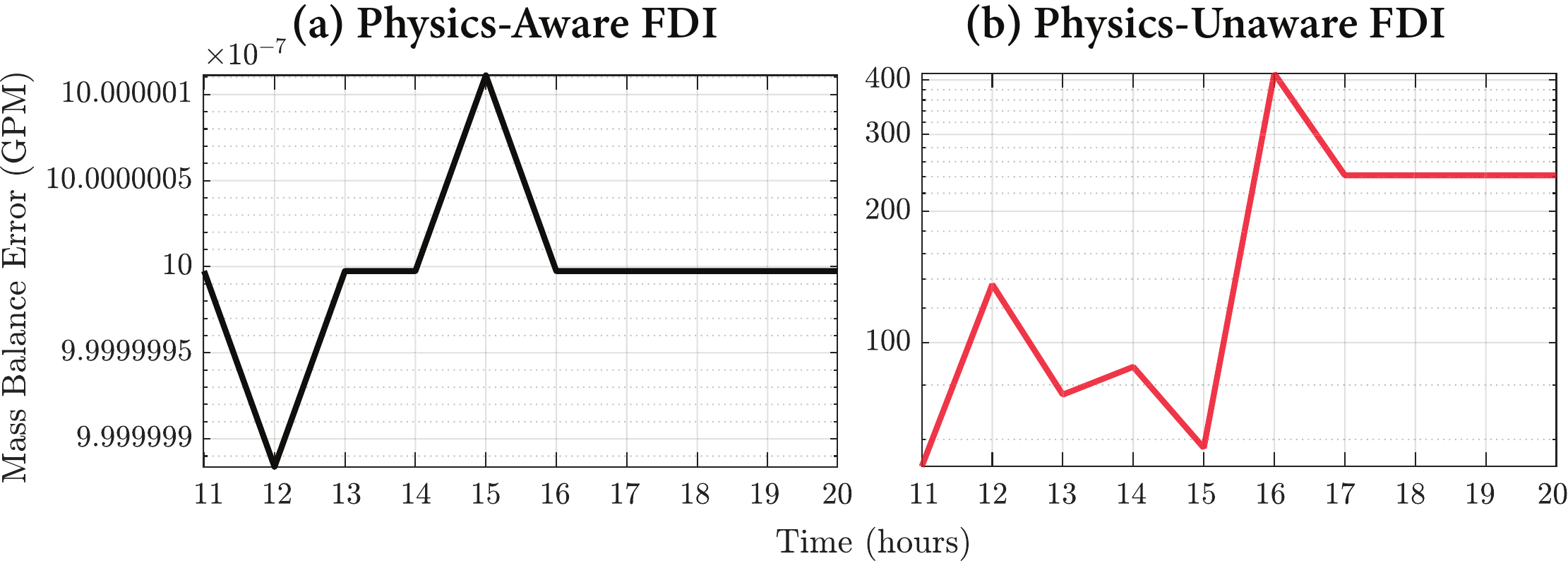}
\caption{Mass Balance Error Validation during: (a) Hydraulics-Aware FDI attack and (b) Hydraulics-Unaware FDI attack}
\label{fig:mass_violation_PA}
\end{figure}

Fig.~\ref{fig:mass_violation_PA} demonstrates how Hydraulics-Unaware attacks that ignore hydraulic relationships lead to obvious physical inconsistencies. While HA-FDI maintains mass conservation with errors below $10^{-7}$ GPM, HU-FDI causes violations up to 400 GPM. These discrepancies can be immediately flagged by basic physical validation checks, rendering such attacks ineffective in practice. This explains why attackers are likely to prioritize physical constraint satisfaction, even if it limits their ability to maximize operational disruption.

\begin{figure}[http]
\centering
\includegraphics[width=\columnwidth]{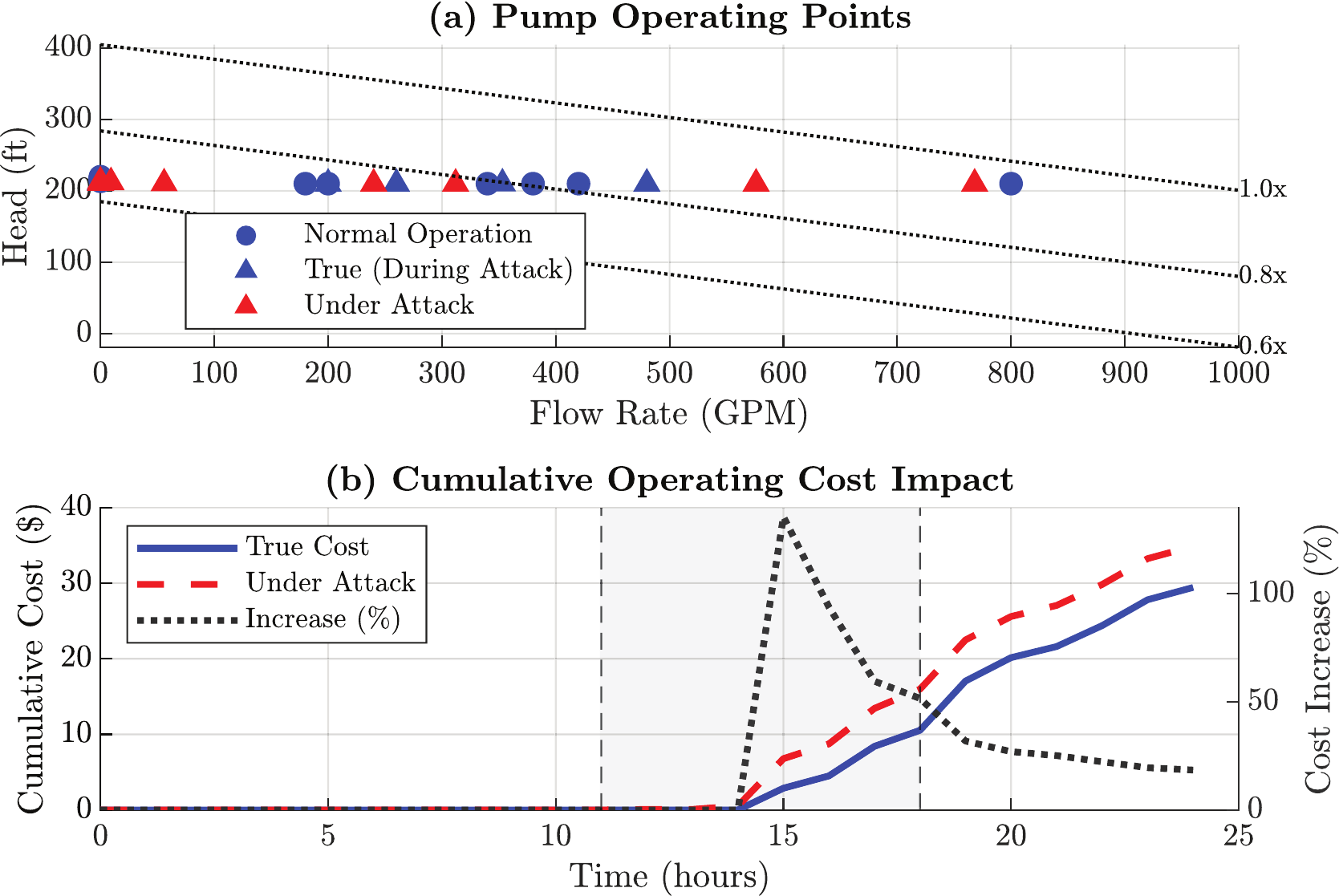}
\caption{Impact of Random FDI on pump operations and cumulative costs}
\label{fig:cost_pump_R}
\end{figure}

The Random FDI strategy represents a simplistic yet potentially the most common attack scenario, as it requires minimal system knowledge beyond access to measurements. Fig.~\ref{fig:cost_pump_R} illustrates its operational impact and shows how the attack forces pump operating points outside optimal regions while remaining within manufacturer-specified curves. A significant increase in the cumulative cost is noticed during the duration of the attack. While these attacks can be easily detected through validation checks and intrusion detection mechanisms, they remain a significant concern for utilities that have not yet implemented such systems. This is particularly concerning as these attacks require minimal sophistication to execute, making them potentially more common in practice.

\subsection{Comparison of Attack Strategies}
\begin{figure}[http]
\centering
\includegraphics[width=\columnwidth]{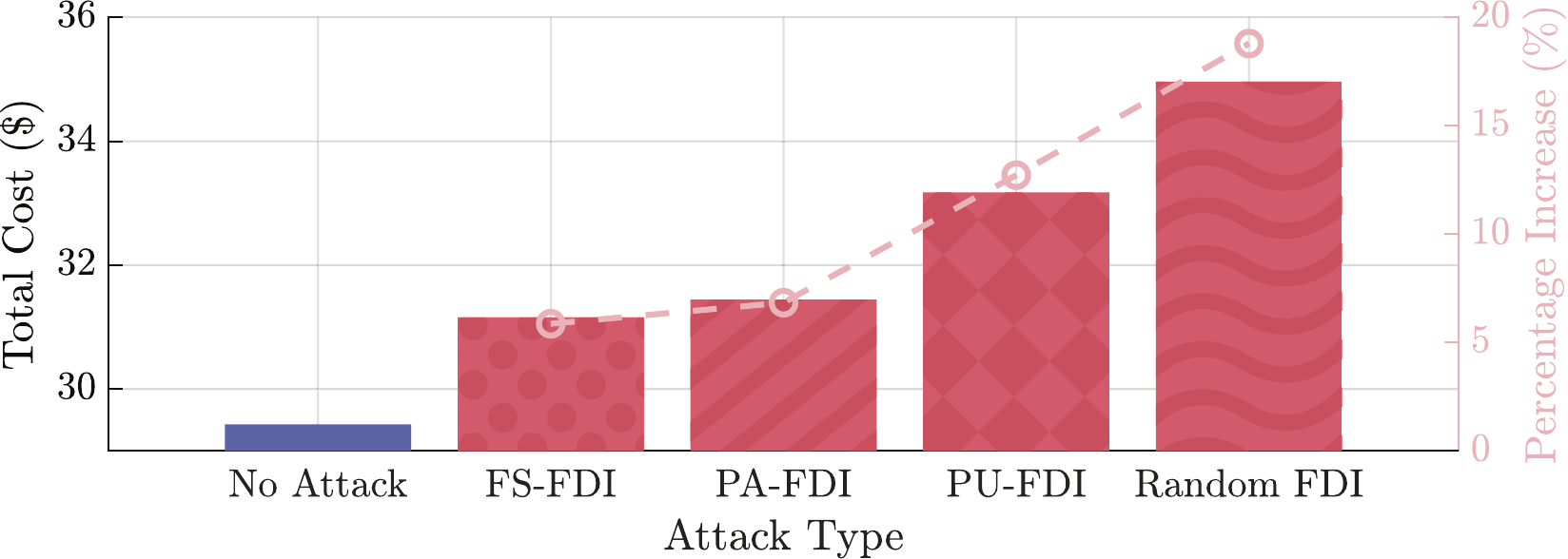}
\caption{Comparison of attack strategies' impact on total operational cost}
\label{fig:comparison}
\end{figure}

The comparative analysis in Fig.~\ref{fig:comparison} demonstrates the fundamental trade-off between attack impact and stealth.

The results show a clear progression in attack impact: FS-FDI and HA-FDI achieve similar cost increases while maintaining physical plausibility, HU-FDI shows higher impact but fails physical validation, and Random FDI causes the largest disruption but would be easily detected through both physical and statistical checks. This hierarchy demonstrates how real-world constraints fundamentally limit the ability of FDI attacks to maximize system disruption while maintaining stealthiness, while also highlighting the potential vulnerability of systems lacking detection mechanisms to even simple attacks.

While these results are from a benchmark network, their implications for real-world systems can be significant. In metropolitan water networks where daily pumping costs reach hundreds of thousands of dollars, sustaining a 5\% cost increase through stealthy attacks could accumulate to millions in excess operational costs annually as these attacks can persist undetected for extended periods of time.

\subsection{Impact on Water Flow Operations}
While previous analysis focused on attacks targeting pump control optimization, the proposed strategies are generalizable and can be adapted for other operational objectives.	

\begin{figure}[http]
\centering
\includegraphics[width=\columnwidth]{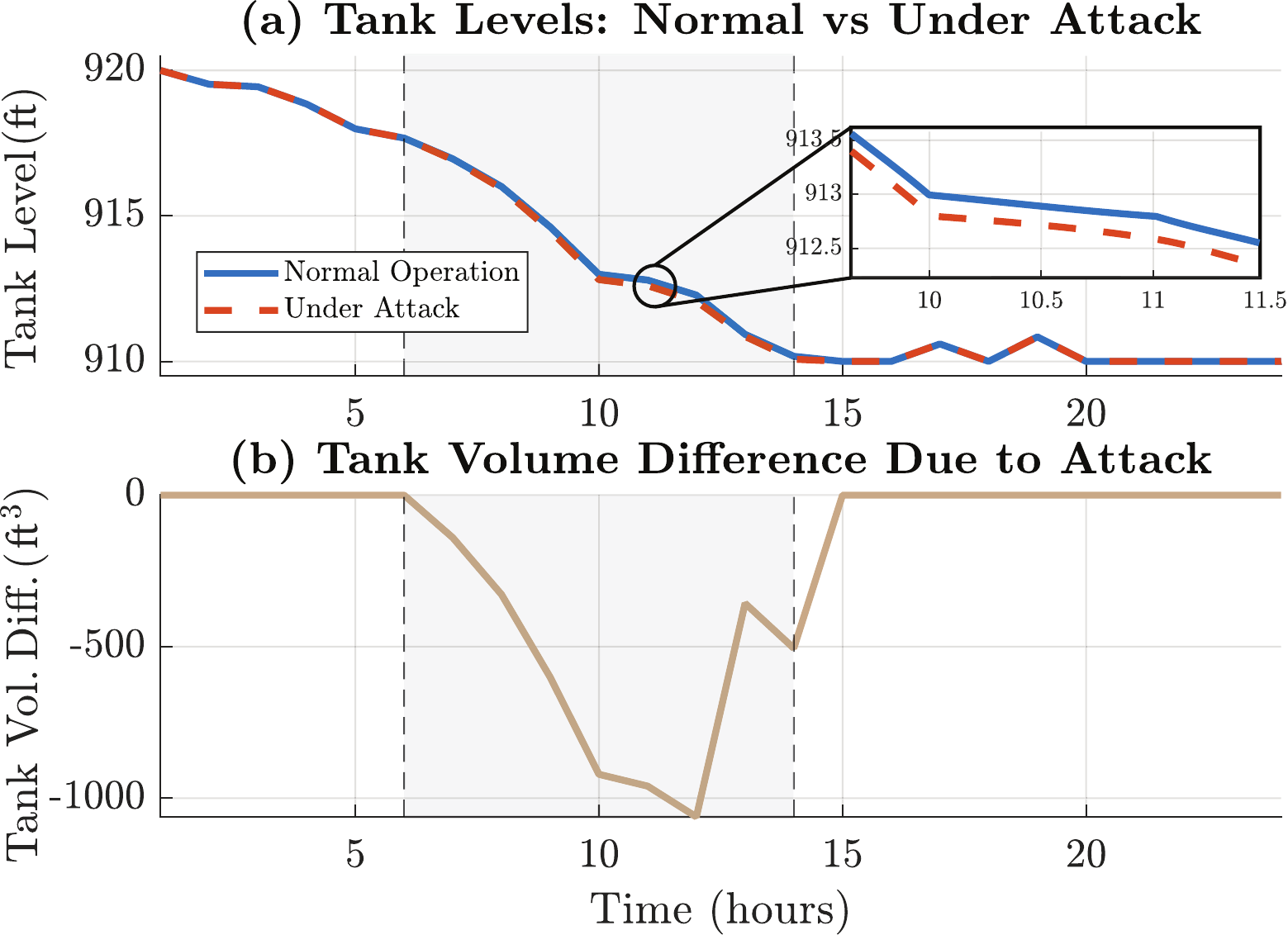}
\caption{Random FDI impact on tank water levels and cumulative volume loss}
\label{fig:waterflow_tank}
\end{figure}

The attack, executed between $t=6$ and $t=14$, manipulates measurements to induce excessive water withdrawal from storage tanks. Fig.~\ref{fig:waterflow_tank} demonstrates how a simple Random FDI attack can effectively target water flow management processes. It shows how subtle measurement modifications lead to cumulative tank level reductions. Fig.~\ref{fig:waterflow_tank}(b) quantifies the total volume impact, an additional 4,874 ft$^3$ drawn from the tank during the 8-hour attack period. This illustrates how the proposed attack strategies can be generalized beyond cost manipulation, though their effectiveness depends on the attacker's measurement access and the objective they aim to achieve.

The ability to construct effective attacks requires careful selection of target measurements based on both accessibility and hydraulic relationships. While the optimization framework can often find feasible solutions that satisfy physical and detection constraints, achieving specific operational impacts requires strategic targeting of measurements that can influence the desired control variables. However, our simulations demonstrate that even random selection of flow and demand measurements typically yields detrimental effects on system operation, suggesting that the inherent coupling of hydraulic variables makes any measurement manipulation potentially harmful. This highlights the importance of understanding system topology when analyzing vulnerabilities and the need for comprehensive monitoring and detection systems in WDNs.

\subsection{Evaluation of Intrusion Detection Methods}
\begin{figure}[http]
\centering
\includegraphics[width=\columnwidth]{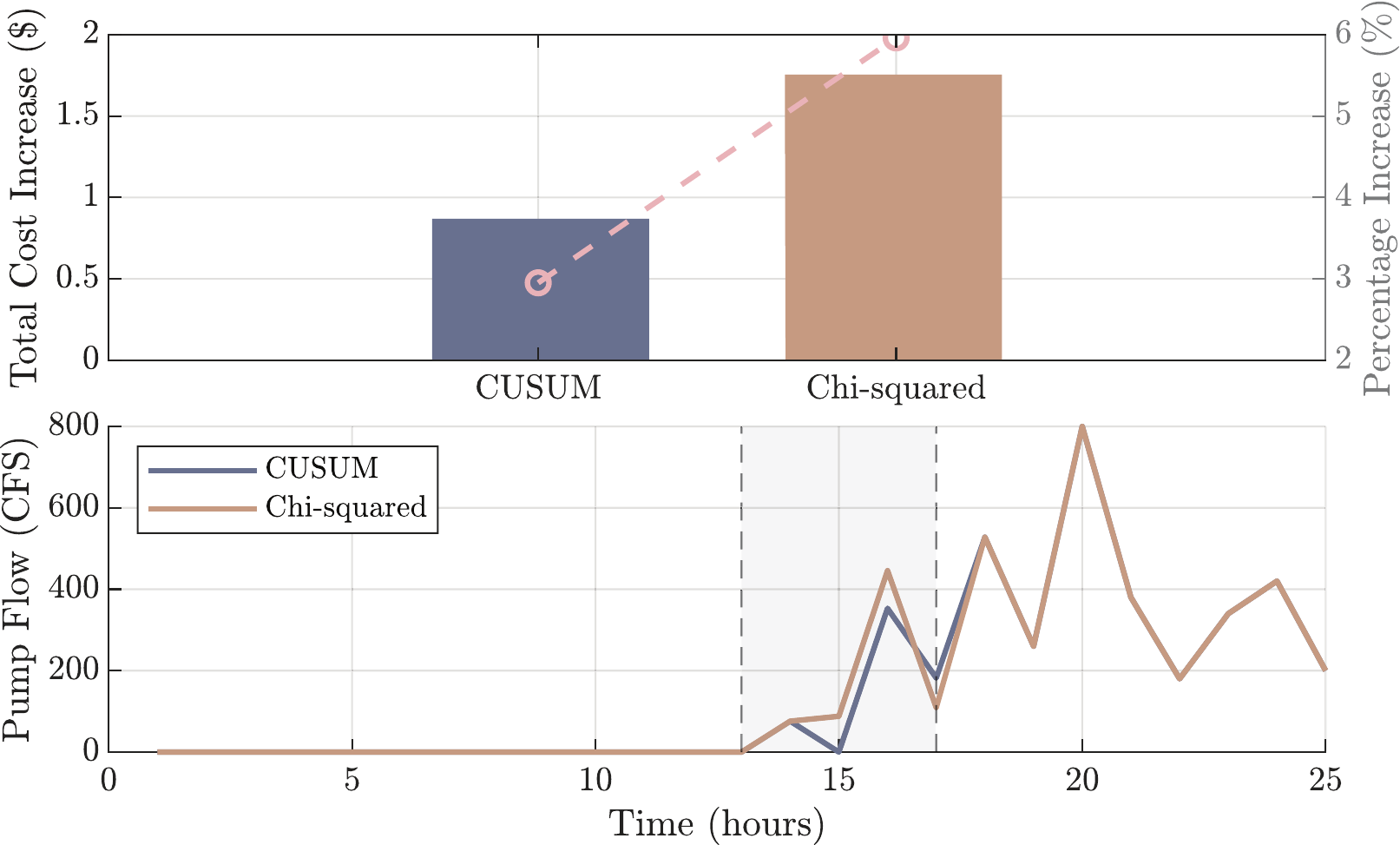}
\caption{Comparison of attack impact under Chi-squared and CUSUM detection}
\label{fig:Chi_vs_CUSUM}
\end{figure}

To quantify the constraints posed by intrusion detection on attack effectiveness, we compare two commonly employed ID methods described in Section~\ref{subsec:ID}: a static Chi-squared detector that uses a scalar distance measure to collectively monitor residuals, and a vectorized dynamic CUSUM detection that monitors each measurement independently. Fig.~\ref{fig:Chi_vs_CUSUM} shows the comparative performance of these detectors against an FS-FDI attack targeting pump flow sensor and Junction 1 demand between $t=13$ and $t=17$ hours.

The results demonstrate that the Chi-squared detector allows for larger measurement manipulations without triggering alarms as seen by the higher pump flow deviations in the Fig.~\ref{fig:Chi_vs_CUSUM}(b). This behavior aligns with previous findings in the literature \cite{ahmedModelbasedAttackDetection2017}. The increased attack tolerance under Chi-squared detection stems from its aggregation of residuals, which can mask localized anomalies, while CUSUM's dynamic accumulation of deviations enables better detection of persistent changes in individual measurements. This also suggests that vectorized detection approaches may provide better protection against tailored attacks, though at the cost of increased false alarm rates and more complex implementation due to the need for individual threshold tuning and historical data collection for each measurement point.

The trade-off between detection sensitivity and false alarms was examined across different threshold values. While tighter thresholds improve attack detection, they lead to increased false alarms during normal operation. Conversely, relaxed thresholds reduce false alarms but create larger blind spots for attackers to exploit. This inherent compromise persists even with carefully tuned parameters, highlighting the fundamental limitations of scalar detection methods against tailored attacks.

\subsection{Scalability and Implementation on Larger Networks} \label{subsec:net3}
\begin{figure}[t]
\centering
\includegraphics[width=\columnwidth]{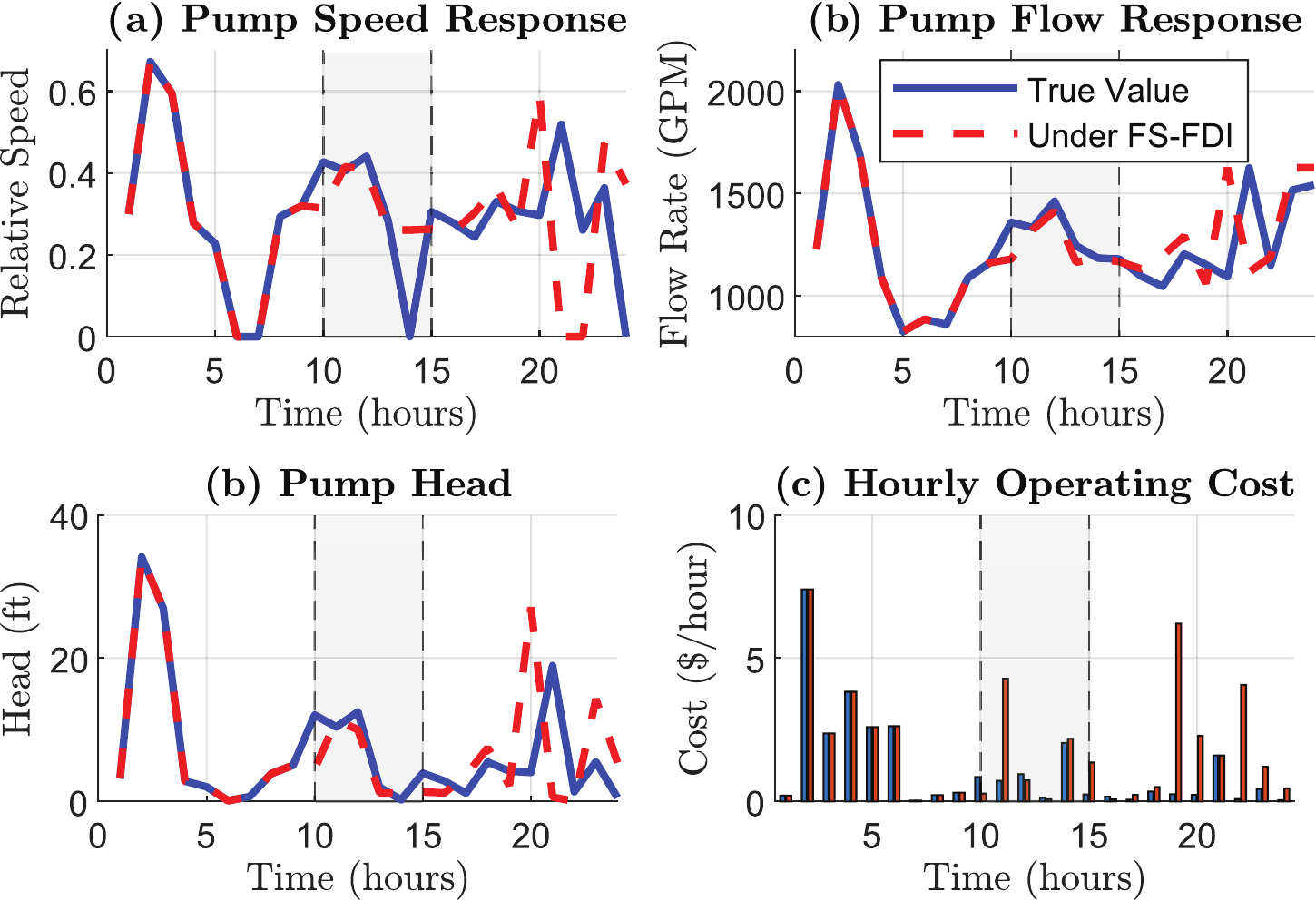}
\caption{Impact of FS-FDI attack on pump operations in Net3 network}
\label{fig:net3}
\end{figure}

The applicability of the proposed attack strategies to larger water distribution networks is primarily constrained by the attacker's access to local system knowledge rather than network size. This is because the designed attacks operate locally---they target specific subsections of the network where the attacker has access to sufficient measurements and system information. When implementing the FS-FDI attack on the larger Net3 network, we observe similar success in manipulating pump operations while maintaining both physical consistency and detection avoidance, despite the network's increased complexity, as seen in Fig.~\ref{fig:net3}.
The results demonstrate how small local perturbations can induce significant system-wide impacts in complex networks. During the attack period (hours 10-15), we implemented coordinated modifications to Pump 1's flow rate measurements and demand readings at its adjacent junction, limited to 5\% of true values. The impact persisted beyond the attack window, resulting in a 62.74\% increase in operational costs over the 24-hour period. This disproportionate impact occurs because the small demand changes force the pump to operate in less efficient regions of its characteristic curve, while the network's interconnected nature means these local inefficiencies cannot be fully compensated by flow from connected pipes. 

The scalability of the proposed attacks stems from the fundamentally local nature of hydraulic relationships in water networks. An attacker with knowledge of local topology, hydraulic parameters, and monitoring system configuration can execute these attacks regardless of the overall network size, as long as they can access a sufficient set of hydraulically coupled measurements within their target subsystem. This ensures both practical and computational scalability: the MILP optimization \eqref{FS-FDI} remains tractable as it only needs to consider constraints relevant to the targeted components and their immediate hydraulic connections, independent of the network dimensions.

\textcolor{black}{We note that in our current implementation, the optimizer is allowed to select any pump speed in the interval \([0,\,s_{\max}]\), which can lead to speeds near zero when the solver finds it beneficial from a purely cost-minimization standpoint. In reality, utilities typically enforce practical bounds to maintain pump efficiency and avoid mechanical issues. As our main focus is to demonstrate how stealthy attacks can exploit measurement data to disrupt hydraulic operations (rather than to replicate exact real-world scheduling), we have left such mechanical limits as optional user-defined parameters in the model.}

\section{Systematic Approach to Attack Design and Execution} \label{sec:attack_implementation}
This section presents a methodology for designing and implementing stealthy attacks against water distribution networks. We first discuss strategic measurement selection based on network topology, then present a structured attack implementation workflow, and finally examine practical considerations for real-world deployment.

\subsection{Strategic Measurement Selection for Attacks}
The effectiveness of stealthy attacks depends critically on selecting appropriate measurement targets that satisfy both hydraulic and detection constraints. Our analysis revealed specific network configurations that are particularly vulnerable to manipulation with minimal sensor access:

Four distinct configurations emerged as prime targets for physically-consistent attacks:

\begin{itemize}[leftmargin=*]
\item \textbf{Boundary Nodes with Demand}: Terminal junctions with a single inflow pipe and demand measurement allow attackers to create physically-consistent false states by manipulating just two measurements. In Net1, Junction 1 exemplifies this vulnerability.

\item \textbf{Source-Consumer Paths}: Direct hydraulic paths from reservoirs or tanks to demand junctions through pumps or valves. Manipulating flow at the source and demand at the destination creates a physically consistent attack while maximizing energy costs.

\item \textbf{Storage Elements}: Tanks and their connected pipes represent natural "buffer zones" where flow imbalances can be attributed to level changes. In Net3, manipulating tank levels and adjacent pipe flow readings created consistent false hydraulic states.

\item \textbf{Flow Distribution Points}: Junctions with multiple connecting pipes require manipulating $n-1$ flow measurements to maintain mass conservation, making these targets feasible only when sufficient measurement access exists.
\end{itemize}

The network topology significantly affects vulnerability and attack complexity. The fundamental differences between network structures create distinct attack strategies:

\noindent \textbf{Radial (Tree-like) Networks:} Attacks can be highly localized as flow paths are unique between any two points. This topology offers attackers several advantages: \textit{(i)} manipulations remain confined to downstream components, \textit{(ii)} mass conservation requires modifying fewer measurements, and \textit{(iii)} the unique flow paths simplify physical consistency constraints. In Net1, the connection from the reservoir through pump to junction 1 exemplifies this vulnerability.

\noindent \textbf{Looped Networks:} The redundant flow paths create hydraulic interdependencies that require more sophisticated approaches: \textit{(i)} Full Loop Manipulation---modify all measurements except one around a complete loop, effectively pushing any physical inconsistency to an unmeasured location; \textit{(ii)} Demand Absorption---target loops containing demand junctions where flow imbalances can be attributed to demand variations; \textit{(iii)} Loop-Breaking---focus on hydraulic control points that enforce directional flow, reducing the problem to a simpler radial-like scenario. Our analysis of Net3 demonstrated how attackers exploiting these properties could maintain physical consistency with minimal sensor manipulations rather than controlling all measurements in a loop.

\begin{figure}[t!]
\centering
\includegraphics[width=\columnwidth]{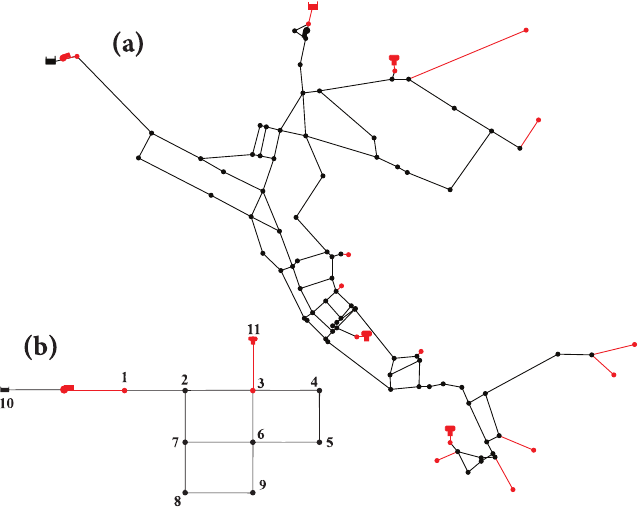}
\caption{\textcolor{black}{Potential targets for physically-consistent attacks in water distribution networks. Highlighted red components indicate strategic attack points requiring minimal sensor manipulations}}\label{fig:targets}
\end{figure}

As illustrated in Fig.~\ref{fig:targets}, both Net3 (a) and Net1 (b) contain multiple potential attack targets (highlighted in red) that require minimal sensor manipulations for the purpose of physically-consistent attacks (such as FS-FDI and HA-FDI attacks). These include terminal demand nodes, pump-reservoir connections, and tank interfaces where attackers can create physically consistent false states with access to just 2-3 measurements.
\subsection{Attack Implementation Workflow}
To facilitate systematic attack simulation and vulnerability assessment, we developed a comprehensive workflow that guides users from measurement selection to attack execution. This process is formalized in Algorithm~\ref{alg:general_attack_simulation} (see \ref{appndx:algo}), which outlines the complete procedure for implementing any of the four FDI attack strategies based on available system knowledge and access privileges.

The algorithm provides a structured approach for practitioners to evaluate vulnerabilities without requiring specialized expertise in optimization theory or hydraulic modeling. It includes crucial feasibility checks, constraint validation, and decision points that ensure generated attacks remain within realistic operational parameters. When constraints cannot be satisfied, the framework suggests alternative sensor groupings or parameter adjustments to identify feasible attack scenarios.

To illustrate this workflow, we present a representative attack scenario based on our Net1 case study. The process begins with strategic target selection: the pump flow sensor and Junction 1 demand meter were identified as ideal candidates due to their hydraulic connectivity and operational significance. These components meet the "boundary node with demand" vulnerability pattern described earlier, requiring only two measurement manipulations to maintain physical consistency.

After target selection, the framework performs hydraulic connectivity validation to ensure the selected measurements can be manipulated in a physically consistent manner. For our Net1 example, the mass balance relationship $(Pump\ flow + a^p_k) - Pipe\ 1\ flow = (Demand + a^d_k)$ must be preserved. The optimization then computes attack vectors that maximize operational impact while satisfying both physical constraints and detection thresholds.

The attack execution, as demonstrated in Fig.~\ref{fig:FS}, shows how coordinated manipulations during hours 15-20 increased apparent pump flow and demand, inducing a 7-15\% cost increase while maintaining physical plausibility and evading detection. This pattern scales effectively to larger networks, as confirmed in our Net3 implementation (Fig.~\ref{fig:net3}), where similar principles generated a 62.74\% cost increase using the same algorithmic approach.

This workflow demonstrates that effective attacks do not require comprehensive network knowledge, but rather targeted understanding of critical hydraulic relationships in specific subsystems. The adaptability of our framework allows for systematic vulnerability assessment across diverse network configurations, providing utilities with practical tools to evaluate security weaknesses and develop appropriate countermeasures.
\color{black}

\subsection{Implementation Considerations and Practical Aspects}
The practical deployment of these attack strategies depends heavily on both the network topology and the attacker's access to system information. Here we discuss key implementation aspects focused on measurement selection and real-world execution constraints.
\textcolor{black}{\subsubsection{State Estimation Robustness}
The effectiveness of state estimation in real-world WDNs is influenced by measurement noise characteristics, sensor placement density, and model parameter uncertainty. Our implementation accounts for measurement noise through the diagonal weighting matrix $\m{W}_m$, assigning lower weights to sensors with higher variance. While sparse sensor networks---common in water utilities due to installation costs---increase vulnerability to targeted manipulations, they also constrain attackers by limiting the available measurement points for coordinated attacks. Robustness can be improved through strategic redundant sensor placement at hydraulic bottlenecks and critical control points. Additionally, model parameter uncertainty (e.g., in pipe roughness coefficients and demand patterns) introduces natural variability in residuals that detection systems must accommodate, potentially creating margins that sophisticated attackers can exploit while remaining below alarm thresholds.}

\subsubsection{Implementation and Technical Reproducibility}
The implementation of these strategies requires compromising field devices like RTUs and PLCs that interface with physical sensors through industrial protocols \cite{Rajabpour2015}. The required system knowledge varies by attack type, while random manipulation may succeed with basic network access, sophisticated attacks require understanding of both system topology and operational patterns, typically gathered through extended network reconnaissance and SCADA traffic monitoring \cite{Torrisi2014Peekaboo}.

\textcolor{black}{The attacks are implementable through standard optimization techniques with computational requirements that remain tractable even for large networks. Specifically, the resulting Mixed-Integer Linear Programs (MILPs) scale primarily with the number of targeted sensors and their immediate hydraulic connections rather than with overall network dimensionality. For typical attack configurations targeting 2-3 sensors in a subsystem, the optimization involves approximately $O(10-20)$ binary variables for the piecewise linearization. In our computational experiments using Gurobi on standard hardware, these MILPs consistently solve within 0.5-2 seconds even for the larger Net3 network—orders of magnitude faster than typical hydraulic time steps (15-60 minutes) in operational WDNs. This computational efficiency combined with the relatively slow WDN dynamics provides ample time for execution in real-world scenarios.} Implementation can be achieved through either hydraulic simulators like EPANET \cite{rossman2020epanet} or custom optimization frameworks. While FS-FDI attacks require significant preparation and system knowledge, they represent worst-case scenarios for security evaluation. Given that many utilities lack proper monitoring systems, simpler attacks requiring minimal knowledge may pose more immediate practical threats, highlighting the importance of implementing basic security measures like state estimation and intrusion detection.

\subsubsection{Security Recommendations}
Based on the analysis of attack strategies and their operational impacts, we propose several recommendations for water utilities. Sensor placement should prioritize hydraulically coupled measurements to enable cross-validation with critical subsystems like pump stations and storage tanks warranting redundant monitoring through independent sensor types. This enables detection of local inconsistencies while limiting an attacker's ability to coordinate measurement manipulations.

Detection system implementation should prioritize physical validation checks across hydraulically connected components, as these provide fundamental protection regardless of attack \textit{sophistication}. This can be enhanced with statistical monitoring and state estimation when resources permit. From an operational perspective, SCADA system segmentation should account for hydraulic relationships while ensuring control decisions are verified using multiple independent measurements.

\section{Paper Limitations and Future Work} \label{sec:paper_limitations}
While this paper presents a systematic framework for analyzing stealthy false data injection attacks against water distribution networks, several limitations of our current approach should be acknowledged:
\begin{itemize}[leftmargin=*]
\item \textbf{Deterministic Hydraulic Model.} Our formulation relies primarily on a deterministic hydraulic model with measurement noise as the only source of uncertainty, without fully accounting for epistemic uncertainty in network parameters.
\item \textbf{Limited State Estimation Methods.} We employ WLS for state estimation, which represents a static estimation approach that does not leverage temporal correlations or dynamic system behavior that more advanced estimators might capture.
\item \textbf{Localized Attack Scope.} The proposed attack strategies primarily target localized subsystems rather than investigating coordinated attacks against multiple network segments simultaneously.
\item \textbf{Network Scale Limitations.}  The case studies utilize benchmark networks (Net1 and Net3) with limited size and complexity compared to real metropolitan water distribution systems.
\end{itemize}
These limitations reflect the inherent trade-offs in modeling complex cyber-physical systems, where increased model fidelity must be balanced against computational tractability and analytical clarity. Despite these constraints, our approach successfully demonstrates the vulnerabilities present in water distribution networks and establishes a foundation for more comprehensive security assessment.

Future research should extend this work in several directions: incorporating demand and parameter uncertainties through robust or chance-constrained optimization; implementing advanced state estimation techniques such as Extended Kalman Filters for time-series-based analysis; developing specialized detection algorithms that exploit WDN hydraulic constraints; exploring the integration of distributed observer-based approaches from control theory into water system security frameworks; and experimental validation on larger, operational networks. Particularly promising is the potential application of distributed observer schemes and leader-following control paradigms from the control-theoretic literature to WDN security, which could enable more resilient monitoring architectures while maintaining compatibility with industry-standard tools like EPANET. These extensions would transform vulnerability analysis into practical security improvements for critical water infrastructure, addressing the multi-faceted challenges in protecting these essential systems against tailored and stealthy cyberattacks.
\color{black}

\section{Summary and Concluding Remarks}\label{sec:conc}
This paper presents a systematic analysis of stealthy false data injection attacks against water distribution networks. We propose several attack strategies that can bypass both physical validation and intrusion detection while disrupting critical operations like pump control and water flow management. We demonstrate through simulations how attackers with varying levels of system knowledge can manipulate sensor measurements to degrade operational efficiency and hydraulic performance while maintaining apparent physical consistency. 

The analysis reveals how attackers can leverage local hydraulic relationships to construct feasible attacks without requiring complete network knowledge. This locality principle ensures attack viability even in large networks, while also highlighting vulnerable subsystems that merit enhanced monitoring. The current lack of basic validation mechanisms in many water utilities means even straightforward sensor attacks can cause significant operational disruptions, emphasizing the urgent need for implementing fundamental monitoring systems.

Looking forward, as water utilities transition toward increased automation and connectivity, the frameworks developed here will become increasingly relevant for systematic vulnerability assessment and security design. Future work should focus on developing resilient control strategies and distributed detection methods that leverage spatial and temporal correlations between measurements, while accounting for constraints like model uncertainties and sensor calibration errors.

\balance
\bibliographystyle{elsarticle-harv}
\bibliography{Paper02}

\appendix
\section{Weighted Least Squared State Estimation} \label{appndx:WLS}
\textcolor{black}{The WLS state estimation minimizes a weighted sum of squared residuals between measured and estimated values, while incorporating hydraulic constraints. The state vector $\m{x} \in \mathbb{R}^{n}$ contains all pipe flows and junction heads:
$$\m{x} = \begin{bmatrix} \m{q} \\ \m{h} \end{bmatrix},$$
where $\m{q} \in \mathbb{R}^{n_p}$ represents flows and $\m{h} \in \mathbb{R}^{n_j}$ represents junction heads, with $n = n_p + n_j$.\\
For direct measurements from sensors, we construct a measurement matrix $\m{H}_m \in \mathbb{R}^{m \times n}$ where $m$ is the number of sensors:
$$\m{y} = \m{H}_m\m{x} + \m{v},$$
where $\m{y} \in \mathbb{R}^m$ is the measurement vector and $\m{v} \in \mathbb{R}^m$ is the measurement noise vector.\\
Measurement noise is modeled through a diagonal weighting matrix $\m{W}_m$, where each diagonal element is the inverse of the corresponding measurement variance:
$$\m{W}_m = \text{diag}\left(\frac{1}{\sigma^2_{\text{flow}}}, \ldots, \frac{1}{\sigma^2_{\text{head}}}\right).$$
In our implementation, $\sigma_{\text{flow}} = 0.05$ and $\sigma_{\text{head}} = 0.1$.\\
Mass balance equations at each junction are incorporated as additional constraints:
$$\sum_{i \in \mathcal{I}_j} q_i - \sum_{o \in \mathcal{O}_j} q_o = d_j.$$\\
These constraints are represented through a matrix $\m{H}_{\text{mass}} \in \mathbb{R}^{n_j \times n}$ and vector $\m{z}_{\text{mass}} \in \mathbb{R}^{n_j}$ such that $\m{H}_{\text{mass}}\m{x} = \m{z}_{\text{mass}}$.\\
Energy balance across pipes is integrated through linearized head loss equations:
$h_i - h_j = r_p \cdot q_p$.\\
These constraints form matrix $\m{H}_{\text{energy}} \in \mathbb{R}^{n_e \times n}$ and vector $\m{z}_{\text{energy}} \in \mathbb{R}^{n_e}$ such that $\m{H}_{\text{energy}}\m{x} = \m{z}_{\text{energy}}$.\\
\\
The final combined system is:
$$
\underbrace{
\begin{bmatrix} 
\sqrt{\m{W}_m}\m{H}_m \\
\m{H}_{\text{mass}} \\
\m{H}_{\text{energy}}
\end{bmatrix}
}_{\m{H}}
\m{x} 
=
\underbrace{
\begin{bmatrix} 
\sqrt{\m{W}_m}\m{y} \\
\m{z}_{\text{mass}} \\
\m{z}_{\text{energy}}
\end{bmatrix}
}_{\m{z}}
$$
The solution then is: $\hat{\m{x}} = (\m{H}^T \m{H})^{-1} \m{H}^T \m{z}$, where $\m{H}$ and $\m{z}$ represent the combined system matrix and vector. This represents the configuration of the WLS SE used in this work.}

\section{Comprehensive SFDIA Implementation Framework} \label{appndx:algo}
Algorithm~\ref{alg:general_attack_simulation} provides a complete implementation framework for all four attack strategies presented in this work. It guides users through the entire process from sensor selection to attack execution, including parameter specification and constraint verification. The algorithm includes adaptive validation checks that help identify when attacks are infeasible, suggesting alternative approaches based on the available system knowledge and sensor access.
\begin{algorithm*}[http]
\caption{Comprehensive SFDIA Implementation Framework}
\label{alg:general_attack_simulation}
\begin{algorithmic}[1] 
\Require Network configuration, sensor measurements $\m{y}_k$, target set $\mathcal{T}$
\Ensure Modified measurements $\m{y}_k^a$ for vulnerability assessment
\State \textbf{Input:} WDN configuration, measurements $\m{y}_k$, target set $\mathcal{T}$
\State \textbf{Select:} Attack type (FS-FDI, HA-FDI, HU-FDI, R-FDI)
\State \textbf{Initialize:} $\m{a}^h_k = \m{0}$, $\m{a}^f_k = \m{0}$, $\m{a}^d_k = \m{0}$
\State \textbf{Define:} Attack bounds $\alpha_h, \alpha_f, \alpha_d$ (typically 5-10\% of original values)
\If{Attack type = \textit{Full-Stealth FDI (FS-FDI)}}
    \State \textbf{Input:} SE configuration (matrix $\m{H}$, weights $\m{W}$, convergence bound $\epsilon$)
    \State \textbf{Input:} ID parameters (threshold $\tau$, bias $b$ for CUSUM or $\alpha$ for $\chi^2$)
    \State \textbf{Input:} Local hydraulic parameters (pipe connectivity, resistance coefficients)
    \State \textbf{Validate topology:} Verify hydraulic connectivity among sensors in $\mathcal{T}$
    \If{Validation fails}
        \State \textbf{Output:} "Selected sensors cannot satisfy physical constraints"
        \State \textbf{Suggest:} Alternative sensor groupings; \textbf{Exit}
    \EndIf
    \State Compute residuals: $\m{r}_k = \m{y}_k - \m{h}(\hat{\m{x}}_k)$
    \State Solve optimization in Eq.~(13) for $\m{a}^h_k, \m{a}^f_k, \m{a}^d_k$
    \If{Optimization infeasible}
        \State \textbf{Output:} "No feasible attack within given constraints"; \textbf{Exit}
    \EndIf
\ElsIf{Attack type = \textit{Hydraulics-Aware FDI (HA-FDI)}}
    \State \textbf{Input:} Local hydraulic parameters (pipe connectivity, resistance coefficients)
    \State \textbf{Validate topology:} Verify hydraulic connectivity among sensors in $\mathcal{T}$
    \If{Validation fails}
        \State \textbf{Output:} "Selected sensors cannot satisfy physical constraints"; \textbf{Exit}
    \EndIf
    \State Solve optimization in Eq.~(19) for $\m{a}^h_k, \m{a}^f_k, \m{a}^d_k$
\ElsIf{Attack type = \textit{Hydraulics-Unaware FDI (HU-FDI)}}
    \State \textbf{Input:} SE configuration (matrix $\m{H}$, weights $\m{W}$)
    \State \textbf{Input:} ID parameters (threshold $\tau$, bias $b$ for CUSUM or $\alpha$ for $\chi^2$)
    \State \textbf{Select:} Optimization-based or closed-form solution approach
    \If{Optimization-based}
        \State Compute residuals: $\m{r}_k = \m{y}_k - \m{h}(\hat{\m{x}}_k)$
        \State Solve optimization (FS-FDI formulation without physical constraints)
    \Else
        \State Determine detector type (Vectorized/Scalar CUSUM, Chi-squared)
        \State Generate attack using corresponding equation (16, 17, or 18)
    \EndIf
\ElsIf{Attack type = \textit{Random FDI (R-FDI)}}
    \State \textbf{Input:} Parameters ($\sigma_d, \sigma_n, \alpha_n, \alpha_s, p_s, \alpha_{max}$)
    \State Initialize drift component: $\m{a}_0^{\text{drift}} = \m{0}$
    \For{each sensor $i \in \mathcal{T}$}
        \State Generate drift, noise, and spike components according to Eq.~(20-24)
        \State Combine components and apply magnitude bounds
    \EndFor
\EndIf
\State \textbf{Apply attack:} $\m{y}_k^a = \m{y}_k + \m{a}_k$
\State \textbf{Evaluate:} Process $\m{y}_k^a$ through hydraulic solver to assess impact
\State \textbf{Return:} Modified measurements and projected operational effects
\end{algorithmic}
\end{algorithm*}

\end{document}